# SiMPLISTIC: A Novel Pairwise Potential for Implicit Solvent Lipid Simulations with Single-site Models

Somajit Dey*, Jayashree Saha

*Department of Physics, University of Calcutta, 92, A.P.C. Road, Kolkata-700009, India*

Last Modified: February 25, 2021

## ABSTRACT

Implicit solvent, coarse-grained models with pairwise interactions can access the largest length and time scales in molecular dynamics simulations, owing to the absence of interactions with a huge number of solvent particles, the smaller number of interaction sites in the model molecules, and the lack of fast sub-molecular degrees of freedom. In this paper, we describe a maximally coarse-grained model for lipids in implicit water. The model is called 'SiMPLISTIC', which abbreviates for '*Single-site Model with Pairwise interaction for Lipids in Implicit Solvent with Tuneable Intrinsic Curvature*'. SiMPLISTIC lipids rapidly self-assemble into realistic non-lamellar and lamellar phases such as inverted micelles and bilayers, the spontaneous curvature of the phase being determined by a single free parameter of the model. Model membrane simulations with the lamellar lipids show satisfactory fluid and gel phases with no interdigitation or tilt. The model lipids follow rigid body dynamics suggested by empirical studies, and generate bilayer elastic properties consistent with experiments and other simulations. SiMPLISTIC can also simulate mixtures of lipids that differ in their packing parameter or length, the latter leading to the phenomenon of hydrophobic mismatch driven domain formation. The model has a large scope due to its speed, conceptual and computational simplicity, and versatility. Applications may range from large-scale simulations for academic and industrial research on various lipid-based systems, such as lyotropic liquid crystals, biological and biomimetic membranes, vectors for drug and gene delivery, to fast, lightweight, interactive simulations for gaining insights into self-assembly, lipid polymorphism, biomembrane organization etc.

## KEYWORDS

Coarse-graining; Lipid polymorphism; Implicit-solvent; Molecular dynamics; Self-assembly; Lyotropic liquid crystals

## I. INTRODUCTION

Lipids are amphiphiles, usually consisting of a hydrophilic polar headgroup linked to two hydrophobic hydrocarbon chains (tails)[1]. Amphiphiles, based on their effective molecular shape or packing parameter[2], self-assemble in aqueous dispersions into phases with different morphologies[3]. Lipids, likewise, also show polymorphism in aqueous dispersions[1,4]. The rather unusual single chain lysolipids form micelles[5]. Lipids with saturated acyl chains and those with unsaturated chains but bigger headgroups form lamellar bilayers (gel or fluid)[6,7]. Lipids with smaller headgroups, including cholesterols[1], on the other hand, form various non-lamellar phases with negative curvatures such as inverted or reversed micelles, inverted hexagonal and bicontinuous cubic phases[8]. Non-lamellar phases with positive curvatures, however, are uncommon in lipid polymorphism[1]. By non-lamellar phases in the following, therefore, we shall refer to non-lamellar phases with negative curvatures only.

Both lamellar and non-lamellar phase forming lipids are important in natural biology as well as biomedical and industry applications. The cell membrane, for example, contains both lamellar and non-lamellar lipids[9]. The non-lamellar lipids tune the membrane curvature and are predominantly present in the cytosolic leaflet[10,11]. Non-lamellar lipids also find applications in drug delivery[12], gene therapy[13] and cosmetics formulations[14]. Lamellar lipids are widely studied in the contexts of natural biology[15], biomimetics[16], biomedical[17,18] and cosmetic applications[19].

The physics of lipids in water is multiscale – both spatially and temporally – and involves cooperative phenomena. Such many-body, multiscale physics is best probed using computer simulations. Besides complementing experiments quantitatively, molecular simulations can provide significant qualitative insights into the structure, function and dynamics of lipid aggregates[7]. Simulations can also decide between competing continuum theories or even guide experiments[20].

The level of detail in a molecular simulation model practically determines the scale accessible by the simulation. The higher the model resolution, the smaller is the timescale or system size that can be simulated. Even with the current state-of-the-art in high performance computing (viz. massively parallel processing and GPU acceleration) bridging the nanoscale atomistic dynamics to large-scale phenomena such as self-assembly, domain formation, membrane fusion etc. that span micrometers in length and milliseconds in time, is still a far cry for all-atom classical molecular dynamics (MD) simulations[21]. Consequently, some level of coarse-graining (CG) is needed in the models, which trades the full atomic level resolution of an all-atom approach for a unified atom or a molecular level representation. The various approaches to CG modeling of biomolecular systems and solvated amphiphiles are well-discussed in the literature and the reader is referred to Ref. 22 and 23 for excellent overviews on the topic.

Of the different CG methodologies used for modeling lipids in water, a special class consists of implicit solvent, top-down approaches in which the lipids are modeled as single or multisite molecules interacting via ad hoc potentials that are designed for simplicity and speed of computation. 'Implicit solvent'[a] means water is never modeled explicitly, the solvent effects being incorporated within effective interactions between lipids. Any vacuum or unoccupied space in the simulation box, therefore, should be considered as water (Figure 1). These particle-based models are top-down in the sense that they are supposed to reproduce the mesoscopic and macroscopic properties such as self-assembly, membrane elasticity etc. without considering the microscopic details explicitly. An accessible review of such implicit solvent coarse-grained (ISCG) models can be found in Ref. 24.

The goal of the present paper is to present a novel ISCG model for self-assembling lipids, where the lipids are modeled as single-site directed ellipsoids interacting via a simple, two-body, anisotropic potential. Being molecular level (i.e. single-site) and implicit solvent, the number of site-site interactions is minimal and the MD timestep can be maximum, because of which this model can simulate lipid systems at the largest scales accessible by any particle-based model for given computational resources. The pairwise interaction potential features a tuneable parameter that determines the spontaneous curvature of the self-assembled phase. For ease of reference, we shall henceforth call this model 'SiMPLISTIC', which is an acronym for '*Single-site Model with Pairwise interaction for Lipids in Implicit Solvent with Tuneable Intrinsic Curvature*'.

To put SiMPLISTIC in perspective, let us briefly discuss some of the previous attempts at minimalistic,[b] ISCG modeling of lipids for off-lattice simulations. The pioneering model by Drouffe et al.[25] consisted of directed spheres, which, using a multibody hydrophobic potential, self-assembled into monolayers. To simulate a lipid bilayer with this monolayer, each sphere would be viewed as containing a hydrophobic layer sandwiched between two (top and bottom) hydrophilic layers. This model, therefore, represented a pair of lipids instead of a single lipid. Noguchi and Takasu[26] were the first to give an ISCG model of a single lipid proper. Their model consisted of a rigid, linear trimer of one hydrophilic and two hydrophobic beads. Self-assembly was still due to a multibody hydrophobic potential. Pairwise potentials, however, are far more computationally efficient than multibody ones. This led to the 'Water-free' model by Farago[27] where each lipid was still modeled as a rigid, linear trimer, but now all interactions were pairwise. Brannigan and Brown[28] took a different approach and brought nonspherical components into the picture. Their model consisted of a nematogenic spherocylinder with a hydrophobic site at its one end, all interactions being pairwise. The molecules in the four models mentioned above are essentially rigid linear rotors with only five degrees of freedom. The fluid bilayers produced by them, however, were either too flexible (forming vesicles easily)[25,26] or too stiff (comparable to membranes with cholesterols)[27,28]. To remedy this, Wang and Frenkel[29] introduced flexibility in the bead-based trimer model by using finite extensible nonlinear elastic (FENE) bonds to link the beads. By this time, it had already become clear for the workers in this field that simple Lennard-Jones (LJ) potentials are not suitable for stabilizing a fluid bilayer. To this effect, some had to use density-dependent multibody potentials[25,26,29] and some[28] employed anisotropic, inverse-squared attraction with a longer interaction range. Although Farago[27] was able to stabilize a fluid bilayer with LJ type interactions only, he had to make the pair potentials non-additive and remain limited to only a heavily tuned set of parameters. Farago's strategy required the hydrophobic beads to see the hydrophilic beads as larger than they actually were, thus creating a steric barrier against lipid evaporation from the membrane plane. As expected, this compromised self-assembly and only preassembled bilayers could be studied. Cooke, Kremer and Deserno[30,31] then realized the capability of broad attractive tail potentials to accomplish self-assembly into stable fluid bilayers. Their model[30] was again a flexible (FENE linked) trimer with the hydrophobic beads attracting each other pairwise and the hydrophilic beads providing soft-core, steric interaction only. The range of the pairwise attraction, however, was now a tuneable parameter that affects the thermal stability and the elasticity of the bilayer. For completeness, we must mention the relatively recent model by Noguchi[32], where each lipid is modeled as a directed sphere with two interaction sites and self-assembly is ensured using a multibody hydrophobic potential.

---
[a] Some authors use the term 'Solvent-free' instead.
[b] By 'minimalistic', we mean models with up to three interaction sites only. For other ISCG models see Ref. 51,52,84–86 and the references therein.

A notable commonality between the above-mentioned ISCG models with pairwise interactions is that none of them incorporates the hydration force[33], which is a key feature of hydrated amphiphile aggregates. Hydration force accounts for the effective repulsion between two hydrophilic layers in close contact, and decays exponentially with the distance between them. This 'hydration barrier' is the reason, apart from thermal fluctuations such as protrusions and undulations, why membrane fusion is difficult[34] and why a bilayer cannot sit immediately on top of another in multilamellar stacks[c] (Figure 1a). In 2017, we published a single-site, ISCG model for amphiphiles, with a pairwise, anisotropic potential, that could mimic such a 'hydration barrier'.[35] Amphiphiles were modeled as soft-core directed spheroids that attracted each other, mimicking hydrophobic force, for some relative orientations and repelled each other, mimicking hydration force, for the rest. Because of this, the packing parameter of the molecule could become that of a (truncated) cone even though the steric core was ellipsoidal. The model featured two key parameters: a tuneable range parameter as in the Cooke-Kremer-Deserno model[30], and a packing parameter tuner that would determine the aggregate morphology. Unlike the rigid anisotropic models of Brannigan and Brown[28] and Noguchi[32], our single-site 2017 model required no exclusively hydrophobic or hydrophilic interaction site, and yet showed rapid, unassisted self-assembly into fluid bilayers, rods and micelles. As a further test of this model, we used it to build ISCG models of both rigid and flexible bolaamphiphiles[36], which successfully reproduced many key features of hydrated bolaamphiphile systems in MD simulations.

Although this 2017 model[35] fulfilled our requirement of a computationally fast, conceptually and programmatically simple[d] ISCG model for generic amphiphiles with various packing parameters, it could not generate phases with negative curvatures. Hence, we could not use it to model the non-lamellar lipids. In fact, to the best of our knowledge, none of the above-mentioned minimalistic ISCG models had been shown to produce stable aggregates with negative intrinsic curvatures through spontaneous self-assembly. This is not surprising when we recognize that the formation of inverted morphologies by ISCG models is more involved than merely getting the molecular packing parameters right. To illustrate, an obvious difference between an inverted micelle with its negative curvature and a direct micelle with positive curvature is that the direct micelle does not encapsulate water whereas the inverted micelle does (Figure 1b and c). The hydration barrier due to this captured water should therefore be taken into account appropriately in an implicit solvent model and a simple model that mimics only the packing parameter of the lipids would not suffice[e]. This is where our new ISCG model SiMPLISTIC fits in. Here, the model particles can efficiently self-assemble into inverted micelles that encapsulate vacuum, which is to be interpreted as water because the model is implicit solvent. SiMPLISTIC can also form bilayers. It can therefore model both lamellar and non-lamellar lipids. In addition, for model membranes, SiMPLISTIC can generate elastic properties consistent with experiments and other simulations, unlike the other rigid models discussed <u>above</u>. Other important features implemented in SiMPLISTIC are the hydrophobic mismatch between lipids of unequal lengths for studying demixing and domain formation in multicomponent membranes, mixing rules for studying mixtures of lamellar and non-lamellar lipids, and the ability to form interdigitation-free bilayers regardless of the lipid length. The latter is relevant in view of the fact that phospholipids with symmetric acyl chain lengths do not show interdigitation in bilayers.

SiMPLISTIC, to the best of our knowledge, is the first, truly single-site model for self-assembling lamellar and non-lamellar lipids using a pairwise potential. SiMPLISTIC's ability to model non-lamellar lipids by simply tuning a parameter, however, came as a by-product. Our original scheme was to design the most minimalistic, single-site lipid model for large-scale model membrane simulations with pairwise interactions only. The basic aim, therefore, was to be able to generate fluid, interdigitation-free bilayers through rapid, spontaneous self-assembly, while also taking into account the interbilayer hydration force. The next section (Sec. II) presents the design of SiMPLISTIC with this original purpose in mind. Sec. III then discusses how the ability to self-assemble into aggregates with different curvatures comes along because of our chosen pair potential. Finally, Sec. IV reports the results of model membrane simulations and extends SiMPLISTIC to include hydrophobic mismatch between lipids of different lengths. We conclude this paper in Sec. V with discussions of SiMPLISTIC's speed, scope and future directions.

## II. THE MODEL: SiMPLISTIC

### A. Design Principles

Given the ad hoc nature of any top-down, ISCG model, it is necessary to lay down a set of design principles around which the model would be developed. If certain design principles lead to a successful model, then those principles must have captured some essence of the physics involved at the time and length scales appropriate to the model. The design of SiMPLISTIC is based on the following core principles.

### 1. Absence of sub-molecular degrees of freedom

Using $^{13}C$ NMR $T_1$ relaxation times obtained from experiments, and Brownian and MD simulations that reproduce the same, Pastor et al.[37–39] had arrived at the following picture of lipid dynamics inside a fluid phospholipid (DPPC) bilayer. Individual lipids average themselves into cylindrical shapes through fast internal (trans-gauche isomerization) and axial motions (rotation about the long molecular axis) on a 100 picosecond timescale. On the nanosecond scale, these cylinders "wobble" diffusively in the viscous bilayer environment. Lateral diffusion of the lipids has an even longer timescale due to restrictions at the bilayer/water interface, but not because of the high viscosity within the bilayer. Much of the mesoscale and macroscale phenomena in fluid bilayers therefore would not be sensitive to the sub-molecular degrees of freedom residing in the acyl chain flexibility. With no need to model this chain flexibility, such large-scale phenomena would be most efficiently simulated using a rigid body model for the lipids. This is because, a rigid, molecular-level model can accommodate a much larger timestep in MD simulations than a flexible model with internal coordinates would ever allow for. Other works in the literature have also found the use of rigid models fruitful[27,28,32,35,40].

---

[c] Adding water to an anhydrous multilamellar stack introduces a layer of water between consecutive bilayers which leads to 'swelling'[87].

[d] The model potential was essentially a (switched) Gay-Berne potential[42] with a modified well-depth and a broad (cubic spline) tail.

[e] For example, no non-lamellar, inverted phase such as the inverted hexagonal, micellar or cubic phase was formed by the Cooke-Deserno model, even when the lipid profile was made inverted conical by making the headgroup bead smaller than the hydrophobic beads[88].

The effectively cylindrical shape of the individual lipids (due to averaging over the rapid axial rotations[41], motivates an axisymmetric model. Because the lipids are not inversion symmetric, a rigid, molecular-level model of lipids must also have an inherent directedness that may be described by a unit vector $\hat{\mathbf{u}}$ directed from the tail end to the head end (see inset of Figure 2). Note that such an axisymmetric and directed model has only five physically significant degrees of freedom— viz. the orientation of its unit vector and its centre of mass (c.m) coordinates. The remaining degree of freedom, viz. the rotation about its axis of symmetry, is physically inconsequential.

## 2.  Single-site model with pairwise interaction

A valid single-site model with a simple, anisotropic pair potential is both economic and elegant compared to an equally valid model with multiple interaction sites, if both the models are to be applied at the same length and time scales. The legacy of the Gay-Berne potential (GB) has been particularly influential in this context. GB[42] is a simple, single-site, anisotropic potential that was originally designed to substitute for a rigid tetramer of LJ sites, and has since seen remarkable popularity in modeling mesogenic molecules in liquid crystal research[43–45].

## 3.  Template for the pair potential

Let us take two model particles $i$ and $j$ with c.m coordinates $\mathbf{r}_i$ and $\mathbf{r}_j$ and unit vectors $\hat{\mathbf{u}}_i$ and $\hat{\mathbf{u}}_j$ respectively. The distance between the c.m is denoted by $r = |\mathbf{r}_{ij}|$, where $\mathbf{r}_{ij} = \mathbf{r}_i - \mathbf{r}_j$. The generic form of a soft-core, pair potential between them can now be written in the well-known Weeks-Chandler-Anderson[46,47] format as

$$V_{ij} = V_{ij}^{\text{Repulsive}} + V_{ij}^{\text{Attractive}},\tag{1}$$

where

$$
\begin{aligned}
V_{ij}^{\text{Repulsive}} &= V^{\text{Steric}}(r) \text{ if } r < r_{zf}\\
&= 0, \text{ otherwise}
\end{aligned}\tag{2}
$$

and

$$
\begin{aligned}
V_{ij}^{\text{Attractive}} &= -\varepsilon_{\text{wd}}, \text{ if } r < r_{zf}\\
&= s(r)\varepsilon_{\text{wd}}, \text{ if } r_{zf} \le r < r_{\text{cut}}\\
&= 0, \text{ otherwise}
\end{aligned}\tag{3}
$$

Meanings of the variables used above are the following. $r_{zf}$: $r$ at which the potential is minimum and the particles exert zero force on each other; $r_{\text{cut}}$: $r$ at which the potential decays to zero or is cut-off; $\varepsilon_{\text{wd}}$: the well-depth of the attractive potential; $s(r)$: a spline curve that switches from -1 at $r_{zf}$ to 0 or near zero at $r_{\text{cut}}$; $V^{\text{Steric}}(r)$: A strictly repulsive potential describing the soft-core. Contextually, it might be noted that the popular LJ and GB potentials also comply with the format given in Eq. (1)-(3). As in GB, any anisotropy in the pair potential is introduced through the dependence of $r_{zf}$ and $\varepsilon_{\text{wd}}$ on the relative orientation of the particles. $r_{zf}(\hat{\mathbf{r}}_{ij}, \hat{\mathbf{u}}_i, \hat{\mathbf{u}}_j)$, for example, would correspond to the nonspherical shapes of the molecular cores.

Desirably, the pair potential must be symmetric under swapping of the particles, viz. $V_{ij} = V_{ji}$. For an anisotropic potential, this implies that the functions $\varepsilon_{\text{wd}}(\hat{\mathbf{r}}_{ij}, \hat{\mathbf{u}}_i, \hat{\mathbf{u}}_j)$ and $r_{zf}(\hat{\mathbf{r}}_{ij}, \hat{\mathbf{u}}_i, \hat{\mathbf{u}}_j)$ should depend only on terms symmetric in $i$ and $j$. Some of the simplest examples of such terms are $(\hat{\mathbf{u}}_i \cdot \hat{\mathbf{u}}_j)$, $(\hat{\mathbf{r}}_{ij} \cdot \hat{\mathbf{u}}_i + \hat{\mathbf{r}}_{ij} \cdot \hat{\mathbf{u}}_j)$ and $(\hat{\mathbf{r}}_{ij} \cdot \hat{\mathbf{u}}_i)(\hat{\mathbf{r}}_{ij} \cdot \hat{\mathbf{u}}_j)$. The directedness or inversion asymmetry of the lipids also requires $\varepsilon_{\text{wd}}(\hat{\mathbf{r}}_{ij}, \hat{\mathbf{u}}_i, \hat{\mathbf{u}}_j) \neq \varepsilon_{\text{wd}}(\hat{\mathbf{r}}_{ji}, -\hat{\mathbf{u}}_i, \hat{\mathbf{u}}_j)$.

## 4.  Phenomenology

Being top-down, our ISCG modeling approach needs to be guided by some phenomenology of generic lipid bilayers. A preference for certain relative configurations between a pair of nearest neighbor lipids in a hydrated bilayer can be easily ascertained from the schematic in Figure 1a. The lipids labeled 1 and 2 are in the side-side parallel (SSP) configuration, whereas those labeled 1 and 3 are in the tail-tail collinear (TTC) configuration. In a stable bilayer, lipids 1, 2 and 3 are bound to each other. If all the lipids interact only through an implicit solvent pair potential as described in Sec. II.A.3, this implies that the SSP and TTC configurations must have negative pair potential energies or, equivalently, positive well-depths ($\varepsilon_{\text{wd}}$).

Lipids 1 and 4 in Figure 1a, on the other hand, belong to adjacent bilayers. They, therefore, should repel each other in their head-head collinear (HHC) configuration so as to mimic the effective hydration force between those bilayers. The potential described in Eq. (3), therefore, needs to be repulsive for the HHC configuration, which is possible if $\varepsilon_{\text{wd}}^{\text{HHC}} < 0$.

Figure 2 lists all possible side-side and collinear configurations between a pair of model lipids. For the sake of discussion, we shall call these the canonical configurations. Let us now discuss the remaining two canonical configurations that we did not discuss before. If a lipid in a bilayer leaflet is flipped upside down about its centre of mass, it will give rise to the side-side antiparallel (SSA) and the head-tail collinear (HTC) configurations as shown in Figure 3a. Because such a flip is prohibited, as otherwise it would expose the flipped lipid's tail to the surrounding water, the SSA and HTC configurations must be highly unstable. Lipid pairs in SSA or HTC configurations, therefore, must be unbound, which, like the HHC case above, requires $\varepsilon_{\text{wd}}^{\text{SSA}} < 0$ and $\varepsilon_{\text{wd}}^{\text{HTC}} < 0$. Making the SSA and HTC configurations unfavorable, contributes to bilayer stability in other ways too. To illustrate, note that the SSA configuration also arises in case of extreme interleaflet interdigitation, viz. when the lipids from one leaflet are pushed into the other leaflet, thus thinning the bilayer down to the width of a monolayer (Figure 3b). In order to avoid the unstable SSA configuration, therefore, the leaflets start to push each other away. This way, the leaflets can offer a realistic resistance against interdigitation. On the other hand, when a lipid protrudes out of the bilayer plane or tries to escape from the bilayer, it must pass through a configuration where its tail comes near the head of the neighboring lipids, all the lipids being parallel (Figure 3c). This configuration is very close to HTC and hence, must be unstable. Consequently, the bilayer remains stable against lipid protrusion and evaporation, as desired.

So far, we have decided only on the signs of $\varepsilon_{\text{wd}}$ for the canonical configurations listed in Figure 2 by considering their stability or instability. Thermodynamic reasoning enables us to go further. Let $T_{\text{fluid}}$ denote the physiological temperature at which the model lipid bilayer must remain stable and fluid. Because $r_{\text{cut}}$ is finite, this requires each of the intraleaflet and interleaflet nearest neighbor pairwise binding energies to be comparable in order of magnitude to the thermal energy, viz. $O(\varepsilon_{\text{wd}}^{\text{SSP}}) \approx O(\varepsilon_{\text{wd}}^{\text{TTC}}) \approx O(kT_{\text{fluid}})$, where $k$ denotes the Boltzmann constant. In addition, to ensure stability against thermally induced leaflet interdigitation as well as lipid protrusion and evaporation, and to enforce the interbilayer hydration barrier, we require the three unbound canonical configurations to have sufficient potential energy barrier at the physiological temperature.

Eq. (4) summarizes our conclusions for later reference.

$$\begin{aligned}
&\varepsilon_{wd}^{SSP}, \varepsilon_{wd}^{TTC} > 0, \quad \varepsilon_{wd}^{HHC}, \varepsilon_{wd}^{SSA}, \varepsilon_{wd}^{HTC} < 0 \\
&O\left(\varepsilon_{wd}^{SSP}\right) \approx O\left(\varepsilon_{wd}^{TTC}\right) \approx O\left(kT_{fluid}\right) \\
&O\left(-\varepsilon_{wd}^{HHC}\right), O\left(-\varepsilon_{wd}^{SSA}\right), O\left(-\varepsilon_{wd}^{HTC}\right) \geq O\left(kT_{fluid}\right)
\end{aligned} \qquad (4)$$

## B. Model Definition

In view of Sec. II.A.3, to define SiMPLISTIC completely, we only need to specify the steric profile, viz. $V^{Steric}(r)$ and $r_{zf}(\hat{\mathbf{r}}_{ij}, \hat{\mathbf{u}}_i, \hat{\mathbf{u}}_j)$, the anisotropic well-depth $\varepsilon_{wd}(\hat{\mathbf{r}}_{ij}, \hat{\mathbf{u}}_i, \hat{\mathbf{u}}_j)$, and the tail $s(r)$.

### 1. Steric profile

Mesogenic molecules are frequently modeled as uniaxial, elongated, rigid objects in liquid crystal simulations, and two simplest shapes normally chosen to represent the shape of the molecular cores[45] are– a) the spherocylinder and b) the Berne-Pechukas (BP) spheroid[48].[f] The BP spheroid is a computationally efficient approximation to the actual spheroidal geometry and forms the steric core in the anisotropic GB potential[42]. The spherocylinder is more efficient for Monte-Carlo (MC) simulations, whereas the BP spheroid is better suited for Molecular Dynamics (MD). This is because, although the contact distance (or distance of closest approach) is simpler to compute for the spheroylinder[49], the computation of the gradients of the contact distance as required for force and torque determination in MD is far cheaper for the BP spheroid[50]. Because our research group uses MD exclusively, we chose the BP spheroid for the core of SiMPLISTIC.[g]

For a BP spheroid of width $\sigma_0$ and length $\sigma_e$, the contact distance is given as

$$\begin{aligned}
\sigma_{BP}(\hat{\mathbf{r}}_{ij}, \hat{\mathbf{u}}_i, \hat{\mathbf{u}}_j) &= \sigma_0 \left[ 1 - \frac{\chi}{2} \left\{ \frac{\left(\hat{\mathbf{r}}_{ij} \cdot \hat{\mathbf{u}}_i + \hat{\mathbf{r}}_{ij} \cdot \hat{\mathbf{u}}_j\right)^2}{1 + \chi(\hat{\mathbf{u}}_i \cdot \hat{\mathbf{u}}_j)} + \frac{\left(\hat{\mathbf{r}}_{ij} \cdot \hat{\mathbf{u}}_i - \hat{\mathbf{r}}_{ij} \cdot \hat{\mathbf{u}}_j\right)^2}{1 - \chi(\hat{\mathbf{u}}_i \cdot \hat{\mathbf{u}}_j)} \right\} \right]^{-1/2} \\
&= \sigma_0 \left[ 1 - \left\{ \frac{\chi\left[\left(\hat{\mathbf{r}}_{ij} \cdot \hat{\mathbf{u}}_i\right)^2 + \left(\hat{\mathbf{r}}_{ij} \cdot \hat{\mathbf{u}}_j\right)^2\right] + 2\chi^2\left(\hat{\mathbf{r}}_{ij} \cdot \hat{\mathbf{u}}_i\right)\left(\hat{\mathbf{r}}_{ij} \cdot \hat{\mathbf{u}}_j\right)\left(\hat{\mathbf{u}}_i \cdot \hat{\mathbf{u}}_j\right)}{1 - \chi^2\left(\hat{\mathbf{u}}_i \cdot \hat{\mathbf{u}}_j\right)^2} \right\} \right]^{-1/2}
\end{aligned} \qquad (5)$$

The anisotropy parameter $\chi$ is related to the aspect ratio $\sigma_0 : \sigma_e$ of the spheroid as $\chi = [(\sigma_e/\sigma_0)^2 - 1]/[(\sigma_e/\sigma_0)^2 + 1]$. Following the soft-core, repulsive part of the GB potential[42], let us now define $V^{Steric}(r)$ for SiMPLISTIC as

$$V^{Steric}(r) = 4\varepsilon_0\left(\frac{1}{R^{12}} - \frac{1}{R^6}\right) + \varepsilon_0, \qquad (6)$$

where $R = (r - \sigma_{BP} + \sigma_0)/\sigma_0$ and $\varepsilon_0$ serves as an energy scale for the model interaction strength. $r_{zf}$ for the above-defined $V^{Steric}(r)$ is at $r = (2^{1/6} - 1)\sigma_0 + \sigma_{BP}$.

### 2. The tail of the model potential

Deserno and coworkers clearly demonstrated the efficacy of a broad attractive tail potential for generating fluid bilayers through self-assembly[31]. In our 2017 amphiphile model[35] and our subsequent work on bolas[36], we too showed successful self-assembly, using a simple cubic spline as the broad attractive tail[h]. In the same vein, $s(r)$ for SiMPLISTIC is now defined as

$$s(r) = (r_{cut} - r)^2 (3r_{zf} - 2r - r_{cut})/(r_{cut} - r_{zf})^3. \qquad (7)$$

Note that the difference $r_{cut} - r_{zf}$ determines the range of pairwise attraction for attractive configurations like SSP and TTC and the range of hydration barrier for HHC. This difference is therefore a parameter of SiMPLISTIC which we shall call *range*.

### 3. Well-depth

In view of the discussion in the last paragraph of Sec. II.A.3, we shall try to form $\varepsilon_{wd}(\hat{\mathbf{r}}_{ij}, \hat{\mathbf{u}}_i, \hat{\mathbf{u}}_j)$ using the terms $(\hat{\mathbf{u}}_i \cdot \hat{\mathbf{u}}_j)$, $(\hat{\mathbf{r}}_{ij} \cdot \hat{\mathbf{u}}_i + \hat{\mathbf{r}}_{ji} \cdot \hat{\mathbf{u}}_j)$ and $(\hat{\mathbf{r}}_{ij} \cdot \hat{\mathbf{u}}_i)(\hat{\mathbf{r}}_{ji} \cdot \hat{\mathbf{u}}_j)$. Note that these three terms are also sufficient to distinguish between all the canonical configurations in Figure 2. This is illustrated in Table 1 below.

| Term | SSP | TTC | SSA | HTC | HHC |
|---|---|---|---|---|---|
| $(\hat{\mathbf{u}}_i \cdot \hat{\mathbf{u}}_j)$ | 1 | -1 | -1 | 1 | -1 |
| $(\hat{\mathbf{r}}_{ij} \cdot \hat{\mathbf{u}}_i + \hat{\mathbf{r}}_{ji} \cdot \hat{\mathbf{u}}_j)/2$ | 0 | 1 | 0 | 0 | -1 |
| $(\hat{\mathbf{r}}_{ij} \cdot \hat{\mathbf{u}}_i)(\hat{\mathbf{r}}_{ji} \cdot \hat{\mathbf{u}}_j)$ | 0 | 1 | 0 | -1 | 1 |

Table 1. Any two configurations (columns) differ in at least one term.

Lipids in a lamellar bilayer mostly remain parallel or antiparallel to each other. Such a preference for alignment can be implemented in SiMPLISTIC simply by using a $\left|\hat{\mathbf{u}}_i \cdot \hat{\mathbf{u}}_j\right|^m$ factor in the expression for $\varepsilon_{wd}$. $m$ is a non-negative, floating-point parameter that is included in order to regulate the strength of the aligning torque.

In view of all the above, and noting that we have already introduced an energy scale $\varepsilon_0$ while defining $V^{Steric}(r)$ in Eq. (6), let us now propose the following ansatz:

$$\varepsilon_{wd} = \varepsilon_0\left(\hat{\mathbf{u}}_i \cdot \hat{\mathbf{u}}_j\right)\left|\hat{\mathbf{u}}_i \cdot \hat{\mathbf{u}}_j\right|^m \left[1 + a\left(\hat{\mathbf{r}}_{ij} \cdot \hat{\mathbf{u}}_i\right)\left(\hat{\mathbf{r}}_{ji} \cdot \hat{\mathbf{u}}_j\right)\right]\left[1 - b\frac{\left(\hat{\mathbf{r}}_{ij} \cdot \hat{\mathbf{u}}_i + \hat{\mathbf{r}}_{ji} \cdot \hat{\mathbf{u}}_j\right)}{2}\right]$$

This form of $\varepsilon_{wd}$ is invariant under swapping of $i$ and $j$ but asymmetric under the inversion $\hat{\mathbf{u}}_i \rightarrow -\hat{\mathbf{u}}_i$, as desired (Sec. II.A.3). Also, $\varepsilon_{wd}$ for the SSP and SSA configurations are respectively $\varepsilon_0$ and $-\varepsilon_0$, which complies with the sign criterion given in the first row of Eq. (4). Note that, of the three parameters— $m$, $a$ and $b$, $m$ has no effect on the values of $\varepsilon_{wd}$ for the different canonical configurations because $\left|\hat{\mathbf{u}}_i \cdot \hat{\mathbf{u}}_j\right| = 1$ for each of them. We, therefore, need to tune $a$ and $b$ in order to make $\varepsilon_{wd}$ satisfy the remaining conditions in Eq. (4). It may also be noted that, fixing $a$ and $b$ this way, leaves $m$ as the only free parameter of the model well-depth, which is very much desirable as the number of parameters becomes minimal.

The two unknowns $a$ and $b$ can be solved for if $\varepsilon_{wd}$ is known for any two of the collinear canonical configurations. To this aim, let us choose $\varepsilon_{wd}$ for the configurations TTC and HTC simply as $\varepsilon_0$ and $-\varepsilon_0$ respectively, in compliance with the sign conditions in Eq. (4).

---

[f] A spheroid is an ellipsoid of revolution and hence is uniaxial or axisymmetric.

[g] For a corresponding MC study, a spherocylindrical core should give the same results as what we obtained using the BP spheroid and MD.

[h] Deserno et al.[30] used a computationally expensive trigonometric function for the tail. Our cubic spline was much cheaper but useful.

Solving for $a$ and $b$ now, we get $a = 2$, $b = 4/3$. Table 2 lists the resulting values of $\varepsilon_{wd}$ for all the canonical configurations. As can be readily verified using these values, for $kT_{fluid} \approx O(\varepsilon_0)$, all the conditions given in Eq. (4) are satisfied.

| SSP | TTC | SSA | HTC | HHC |
|-----|-----|-----|-----|-----|
| $\varepsilon_0$ | $\varepsilon_0$ | $-\varepsilon_0$ | $-\varepsilon_0$ | $-7\varepsilon_0$ |

Table 2. Well-depth $\varepsilon_{wd}$ for the canonical configurations in Figure 2.

## C. Model Summary

To summarize, SiMPLISTIC represents lipids in implicit solvent as directed spheroids, which interact using the following potential.

$$V_{ij}(r) = 4\varepsilon_0 \left( \frac{1}{R^{12}} - \frac{1}{R^6} \right) + \varepsilon_0 - \varepsilon_{wd}, \text{ if } r < r_{zf}$$

$$= \varepsilon_{wd} \frac{(r_{cut} - r)^2 (3r_{zf} - 2r - r_{cut})}{range^3}, \text{ if } r_{zf} \le r < r_{cut} = r_{zf} + range$$

$$= 0, \text{ otherwise}$$

where

$$R = (r - \sigma_{BP} + \sigma_0)/\sigma_0 \ ; \ r_{zf} = (2^{1/6} - 1)\sigma_0 + \sigma_{BP} \ ;$$

$$\varepsilon_{wd} = \varepsilon_0 \left( \hat{\mathbf{u}}_i \cdot \hat{\mathbf{u}}_j \right) \left| \hat{\mathbf{u}}_i \cdot \hat{\mathbf{u}}_j \right|^m \left[ 1 + 2 \left( \hat{\mathbf{r}}_{ij} \cdot \hat{\mathbf{u}}_i \right) \left( \hat{\mathbf{r}}_{ji} \cdot \hat{\mathbf{u}}_j \right) \right] \left[ 1 - \frac{2}{3} \left( \hat{\mathbf{r}}_{ij} \cdot \hat{\mathbf{u}}_i + \hat{\mathbf{r}}_{ji} \cdot \hat{\mathbf{u}}_j \right) \right]$$

$$(8)$$

For the SSP and SSA configurations, $V_{ij}(r)$ is illustrated in Figure 4. Apart from the lipid length $\sigma_e$, which only affects the spheroidal contact distance $\sigma_{BP}$ [Eq. (5)], SiMPLISTIC has only two parameters, viz. $range$ and $m$. The $range$ parameter determines the number of interacting pairs of molecules and hence, quite trivially, affects the overall binding energy of the lipid aggregates which, in turn, affects thermal stability and elasticity[30,31]. $m$, on the other hand, tunes the degree of alignment between the interacting lipids and therefore, for sufficiently large values, should guarantee the formation of bilayers where the lipids possess high orientational order. In the next section, we shall test this hypothesis and also examine SiMPLISTIC's self-assembly behavior for smaller values of $m$.

## III. SELF-ASSEMBLY, TUNEABILITY, MIXING RULE

In this section we report self-assembly behavior of SiMPLISTIC (as seen in MD simulations for different values of $m$), analyze it and extend SiMPLISTIC to model new cases based on that analysis.

## A. MD Study

### 1. Protocol

Self-assembly in an implicit solvent system is usually studied under the NVT ensemble for both MD[26,31,51] and MC[28,52] simulations. For our MD simulations, we chose the Nosé-Hoover thermostat[53], with separate temperature control variables coupled to the translational and rotational degrees of freedom for faster equilibration[54]. Thermostatted translational motion was integrated using the leap-frog scheme given in the Appendix of Ref.[55]. For the rotational motion, the integrator described in Ref.[56] was used.

The system consisted of 1600 identical lipids in a cubic simulation box with periodic boundary conditions (PBC). To work in reduced units, each of the three constants— $\sigma_0$, $\varepsilon_0$, and the mass ($\mu$) of each lipid, was taken as unity. The MD timestep was chosen to be $0.0035\sqrt{\mu\sigma_0^2/\varepsilon_0}$. Regarding the lipid length, we note that aspect ratios $\approx 1:3$ have been employed in several previous works on lipid modeling[26,29,30,40]. In view of this, we also chose lipid length $\sigma_e = 3\sigma_0$. To make the rotational timescale comparable to the translational one, the moment of inertia (about the c.m) for each lipid was chosen to be

$4\mu\sigma_0^2$.

Although $range$ was featured as a free parameter of SiMPLISTIC in Sec. II.C, it is easy to see that a desired value of $kT_{fluid}$ restricts the choices for $range$. This is because too small a value of $range$ would lead to insufficient binding in the face of thermal kinetic energy, and too large a value would turn the lipid aggregate into a tightly bound crystal or gel. For convenience, one might want $kT_{fluid} \approx \varepsilon_0$, and therefore we probed different values of $range$ at $kT = \varepsilon_0$. $range = \sigma_0$ and $2\sigma_0$ did not show self-assembly, but $range = 3\sigma_0$ did. Moreover, at this value of $range$ the bilayer also remained fluid (this will be discussed later in the context of bilayer characteristics, in Sec. IV.A). Later we found that $range = 2.5\sigma_0$ also gives stable phases at $kT = \varepsilon_0$. We, however, performed simulations mostly with $range = 3\sigma_0$ because it leads to faster self-assembly and gives the hydration force a range comparable to the lipid length.

In order to study spontaneous, free self-assembly from a disordered, isotropic phase under NVT, choosing an appropriate number density is very important. Implicit solvent systems are far less dense than anhydrous systems such as systems of mesogens, simply because there is an implicit presence of water in the form of unoccupied volume within the simulation box. Too low a number density, however, is not conducive to self-assembly as can be verified easily. One way to get a viable number density for ISCG lipid systems is to start with an approximately valid preassembled structure such as a system of well-separated bilayers. The disordered, isotropic configuration can then be generated simply by evaporating the preassembled structure at a high temperature under NVT. This way of producing a randomized initialization also eliminates any undesirable core-core overlap between the lipids[28]. As depicted in Figure 5, we generated our disordered, isotropic configuration by evaporating two lattice membranes that were assembled on square grids. To ensure enough space in the simulation box, the two bilayers were kept out of each other's range of interaction, viz. the interbilayer separation was made greater than $range$. This resulted in a box size of $21.3\sigma_0$ and number density of $0.166\sigma_0^{-3}$. All lipids in the disordered phase were then allocated random velocities from the Maxwell-Boltzmann distribution at $kT = \varepsilon_0$ and self-assembly was studied for that temperature for different values of $m$. We infer that the lipids were free to arrange themselves during self-assembly because the overall internal pressure or stress, as computed using the virial[57], remained near zero.

### 2. Self-assembly

Self-assembly was studied for $m = 0, 0.5, 1, ..., 4$ at $kT = \varepsilon_0$. Figure 6 shows some of the phases obtained. Self-assembly was rapid with structure being discernible by 4000 steps and prominent aggregate formation by 6000 steps. For $0 \le m \le 3$, closely packed inverted micelles were obtained, with the micellar curvature decreasing with increasing $m$. For $m > 3$, a preference for bilayer formation was clearly visible, with almost no negatively curved phase remaining in the simulation box at $m = 4$. The parameter, $m$, therefore, determines whether the self-assembled phase will be non-lamellar or lamellar. We also studied self-assembly for some random values of $m > 4$ and only lamellar bilayers were obtained. To summarize our observations, the maximal curvature for non-lamellar phase is obtained for $m = 0$, the curvature of non-lamellar phase decreases with increasing $m$, and no intrinsic curvature, negative or positive, remains in the system for $m \ge 4$.

### B. How $m$ tunes curvature

To explain the aforementioned observations, let us analyze how $m$ affects the energy of lipid splay. We define splay as any configuration between a pair of lipids $i$ and $j$ where each lipid is a mirror image of the other with respect to a mirror plane that bisects and is normal to $\mathbf{r}_{ij}$. The angle between the lipids is called the splay angle, $\theta$ (Figure 7). Note that $\theta = 0, \pi, -\pi$ lead to the SSP, TTC and HHC configurations respectively (cf. Figure 2). As seen from the plot in Figure 7, the most preferred splay angle $\theta_{max}$, viz. $\theta$ at which $\varepsilon_{wd}$ is maximum, is negative for each $m \geq 0$. Also, the amount of splay in this configuration, viz. $|\theta_{max}|$, decreases with increasing $m$. Notice that, because the model lipids are cylindrically symmetric, a preference for non-zero splay implies a conical molecular profile. If the most preferred splay angle is negative, viz. $\theta_{max} < 0$, the effective molecular shape must be that of an inverted cone with the angle between the cone axis and cone surface given by $\theta_{max}/2$. Inverted cones form non-lamellar phases with negative intrinsic curvatures such as the inverted micelles. As $\theta_{max} \to 0$, the effective molecular shape becomes cylindrical which favors the formation of lamellar bilayers with zero spontaneous curvature. From this perspective, it is clear how $\theta_{max}$ governs the spontaneous curvature of the self-assembled shapes. Because $\theta_{max}$, in turn, is tuned by $m$ as discussed above (Figure 7), this explains the observations reported in Sec. III.A.2. Note that this also shows how, despite having a fixed spheroidal core, SiMPLISTIC lipids can have their effective shape or packing parameter[58] varied with $m$ and thus exhibit polymorphism on self-assembly.

## C. Mixing Rule for Lipids with Different Values of $m$

A logical extension of the discussion in the last subsection (Sec. III.B) is to ask how to model the interaction between a pair of SiMPLISTIC lipids each of which generates a different spontaneous curvature in single-component self-assembly. In other words, we seek to extend SiMPLISTIC so that mixtures of lipids with different packing parameters, such as lamellar and non-lamellar lipids, can be simulated.

To this end, let us take two types of SiMPLISTIC lipids, A and B, which differ only in the parameter $m$. From the geometry of cones it follows, $\theta_{max}^{AB} = \left( \theta_{max}^{AA} + \theta_{max}^{BB} \right)/2$. This implies that $m$ for the A-B interaction must be intermediate between that for the A-A and B-B interactions. In keeping with this and the widely adopted Lorentz-Berthelot mixing rule, we propose $\varepsilon_{wd}^{AB} = \sqrt{\varepsilon_{wd}^{AA} \cdot \varepsilon_{wd}^{BB}}$, which gives

$$m_{AB} = \frac{m_{AA} + m_{BB}}{2} . \qquad (9)$$

A two-component mixture of A and B lipids can now be studied with the interactions A-A, B-B and A-B each following Eq. (8) with $m = m_{AA}$, $m_{BB}$ and $m_{AB}$ respectively. To test the validity of this mixing rule, we studied self-assembly for a binary mixture of 800 lamellar and 800 non-lamellar lipids with $m = 4$ and 0 respectively. Figure 8 depicts a few close-ups of the aggregates formed. Note how the lamellar lipids (shown in red) intermix with the non-lamellar lipids (shown in green) in forming the inverted micelle (Figure 8c). The preference of the lamellar lipids to form bilayers is also discernible from the lamellar clusters of the red lipids just beside the inverted micelle (Figure 8b).

## IV. MODEL MEMBRANES, HYDROPHOBIC MISMATCH

This section reports how SiMPLISTIC performs for model membrane simulations. Later in this section, we shall also discuss an extension of SiMPLISTIC to include the effect of hydrophobic mismatch in multicomponent bilayers. Because SiMPLISTIC is required to model only lamellar lipids for membrane simulations, we must take $m \geq 4$ in view of the self-assembly results reported in Sec. III.A.2. All simulations, however, are performed with $m = 4$ only (also see Sec. IV.A.4 for further remarks).

## A. Bilayer Characteristics

Model membrane simulations are usually performed under zero lateral tension[30,32]. For such constant tension simulations, we adopted a two-dimensional Nosé-Hoover NPT dynamics[57] in the bilayer plane to regulate the lateral pressure. To prepare the initial state, a preassembled box-spanning bilayer was thermalized at $kT = \varepsilon_0$ under zero tension. As desired, the surface tension, measured using the diagonal elements of the internal pressure tensor[59], remained near zero[60] for every production run.

### 1. Lateral diffusivity, area per lipid : Fluid and gel phases

The in-plane or lateral diffusivity ($D_\parallel$) of lipids in the membrane is computed from the limiting slope of the mean square displacement (MSD) in the bilayer plane, using the Einstein relation for two-dimensional diffusion[61,62]. Figure 9 depicts the MSDs plotted for four NVE runs starting with equilibrated tensionless bilayers, with different mean temperatures. The corresponding diffusivities and projected area per lipid ($A_L$) are listed in Table 3.

| $kT$ | 0.46 | 0.74 | 0.94 | 1.11 |
|------|------|------|------|------|
| $D_\parallel$ | $1.11 \times 10^{-2}$ | $2.60 \times 10^{-2}$ | $9.94 \times 10^{-2}$ | 0.348 |
| $A_L$ | 0.99 | 1.04 | 1.16 | 1.30 |

Table 3. Lateral diffusivity ($D_\parallel$) and area per lipid ($A_L$) at various temperatures ($kT$). Units: $D_\parallel$ in $\sigma_0 \sqrt{\varepsilon_0/\mu}$, $A_L$ in $\sigma_0^2$, $kT$ in $\varepsilon_0$.

A fluid-gel transition is typically associated with significant drop in $D_\parallel$, reduction of $A_L$ and increased (hexagonal) bond orientation order[28]. From the MSDs plotted in Figure 9 and a visual inspection of the lipid arrangement in the bilayers depicted there, it is apparent that the membrane is fluid at $kT = 1.1 \varepsilon_0$ and $0.94 \varepsilon_0$ and gel at $kT \leq 0.74 \varepsilon_0$. Our objective of generating a fluid phase around $KT_{fluid} \approx \varepsilon_0$ (Sec. III.A.1) is therefore fulfilled.

### 2. Order parameter

The orientational (uniaxial) order parameter for lipids in a bilayer is given as $\frac{1}{2} \left\langle 3 (\hat{\mathbf{u}}_i \cdot \hat{\mathbf{n}})^2 - 1 \right\rangle_i$, where $\hat{\mathbf{n}}$ is a unit vector along the average normal of the bilayer. For a tensionless membrane at $kT_{fluid}$, the order parameter of SiMPLISTIC lipids is 0.967. From this high amount of orientational order, and a visual inspection of the side view of the bilayer as provided in Figure 9, it is safe to infer that the fluid bilayer is not tilted. No tilt was found in the gel phase as well.

### 3. Bilayer thickness

The thickness of a bilayer is estimated as follows. Taking the X and Y axes in the bilayer plane, the bilayer thickness lies along the Z direction. The Z coordinate of the head end of a lipid is given by $h_{\hat{n}} = \left( \mathbf{r} + \frac{\sigma_c}{2} \hat{\mathbf{u}} \right) \cdot \hat{\mathbf{n}}$. The bilayer thickness may be estimated as the spread in the distribution of $h_{\hat{n}}$, which is computed as twice the standard deviation of $h_{\hat{n}}$. For a tensionless fluid membrane, we found the thickness to be $6.297 \sigma_0$. In contrast, for a crystalline flat bilayer with no tilt or interdigitation, the thickness must be $2\sigma_c = 6\sigma_0$. Because our fluid membrane is thicker than this, it implies that there is no interdigitation in the fluid phase, as is also apparent from the side view of the bilayer in Figure 9. No interdigitation was found in the gel phase as well.

### 4. Bending modulus

Being designed with a top-down, phenomenological approach, getting the membrane elasticity such as the bending rigidity right is a decisive factor for SiMPLISTIC. All other top-down attempts at ISCG modeling of lipids with rigid models[25–28] had failed to produce experimentally relevant values for the bending moduli, as was mentioned in Sec. I. There are several methods for determining the bending modulus $\kappa_C$ from configurations generated by molecular simulations[63]. Of these, a method based on spectral analysis of the height fluctuations of a tensionless bilayer has been used by most previous works on ISCG modeling of lipids. This method is difficult and computationally expensive, which has to do with its requirement of very large system sizes (continuum limit)[31] and a robust sampling of global bilayer undulations to apply the Fourier transform on. In contrast, a relatively recent and computationally simpler method[64] based on real-space fluctuations (RSF) of the splay degrees of freedom is local in nature and, therefore, can be applied to rather small system sizes (as small as ~100 lipids), which makes it more practical for MD simulations. This method has also been successfully validated on a range of lipid systems[65]. For our case of flat bilayers composed of SiMPLISTIC lipids with director $\hat{\mathbf{u}}$, the RSF method is briefly described as follows.

According to the RSF methodology[64,66] the splay at any point $\mathbf{p}$ on the bilayer is given as the local covariant derivative of the vector field $\hat{\mathbf{u}} - \hat{\mathbf{n}}$ along any vector $\hat{\mathbf{e}}_\alpha$ tangent to the bilayer. Because the bilayer normal $\hat{\mathbf{n}}$ does not vary across a flat bilayer and $\hat{\mathbf{n}} \cdot \hat{\mathbf{e}}_\alpha = 0$, the splay is given as

$$S_\alpha = \lim_{\Delta \to 0} \frac{[\hat{\mathbf{u}}(\mathbf{p}) - \hat{\mathbf{u}}(\mathbf{p} + \Delta\hat{\mathbf{e}}_\alpha)] \cdot \hat{\mathbf{e}}_\alpha}{\Delta}. \tag{10}$$

Computationally, $S_\alpha$ may be approximated as $\left[(\hat{\mathbf{u}}_i - \hat{\mathbf{u}}_j) \cdot \hat{\mathbf{e}}_\alpha\right] / (\mathbf{r}_{ij} \cdot \hat{\mathbf{e}}_\alpha)$ where $\{i, j\}$ is a pair of nearest neighbor lipids from the same leaflet. The bilayer bending rigidity $\kappa_C$ is related to the frequency distribution of $S_\alpha$ as

$$f(S_\alpha) \propto \exp\left(-\frac{\kappa_C S_\alpha^2 A_L}{2kT}\right). \tag{11}$$

$\kappa_C / kT$ can therefore be determined from the variance of a Gaussian with zero mean fitted to the frequency distribution of $S_\alpha$ (Figure 10). Employing this method, we computed the bending rigidity of a tensionless SiMPLISTIC bilayer at five different temperatures, as listed in Table 4.

| Phase | Fluid | | | Gel | |
|---|---|---|---|---|---|
| $kT$ | 1.20 | 1.11 | 1.00 | 0.80 | 0.46 |
| $\kappa_C / kT$ | 24.6 | 28.3 | 36.9 | 56.5 | 106.9 |
| $\kappa_C$ | 29.5 | 31.4 | 36.9 | 45.2 | 49.2 |

Table 4. Bending modulus ($\kappa_C$) at various temperatures ($kT$). Both $\kappa_C$ and $kT$ are in units of $\varepsilon_0$.

$\kappa_C / kT$ of fluid biological membranes without cholesterol falls within the range $5 - 30$[67,68] Our numerical estimates for fluid SiMPLISTIC bilayers also fall within this range for $kT \geq 1.11\varepsilon_0$ (Table 4). With $kT_{\text{fluid}} = 1.11\varepsilon_0$ or more, SiMPLISTIC therefore successfully reproduces experimentally relevant values for the bending stiffness. This also proves that a rigid, molecular level model of lipids, if designed properly, is capable of generating valid bending moduli for model membranes, despite being stiffer than the flexible bead-spring models (such as Ref. [29,30,52]) that have so far provided the only means to this end.

**Remarks on our choice of $m$**: As mentioned before, all model membrane simulations were performed with $m = 4$. Why choose only

$m = 4$ when $m > 4$ also give lamellar bilayers (Sec. III.A.2)? As shown in Figure 7, increase in $m$ leads to a preference for decreased amount of splay. Because bending stiffness is inversely proportional to the variance of the splay distribution [Eq. (11)], this implies that $\kappa_C$ increases with $m$. With $m = 4$, $\kappa_C / kT$ already lies near the upper end of the experimentally interesting range (Table 4). $m > m_{\text{max}}$ for some $m_{\text{max}} \geq 4$ would therefore lead to unrealistically large values for $\kappa_C / kT$ that lie outside this range. Hence, most values of $m > 4$ are not relevant. One may, however, still probe how big $m$ can get, viz. $m_{\text{max}}$, without generating too big a bending modulus. We leave this exercise for future works.

### 5. Area compressibility modulus and elastic ratio

For constant tension simulations where the membrane area fluctuates about a mean value to regulate the surface tension, the best suited measure for area compressibility modulus $\kappa_A$ is given by[59]

$$\kappa_A = \frac{kT\langle A \rangle}{\langle A^2 \rangle - \langle A \rangle^2} \tag{12}$$

where $A$ denotes the total (projected) area of the membrane and $\langle \ \rangle$ denotes the time average obtained from a production run of the simulation. The elastic ratio is a dimensionless quantity defined as

$$b = \kappa_A h^2 / \kappa_C = \frac{\langle A \rangle h^2}{\langle A^2 \rangle - \langle A \rangle^2} / \left(\frac{\kappa_C}{kT}\right) \tag{13}$$

where $h$ denotes the membrane thickness[28,29]. $b$ ranges from $4 - 12$ for bilayers with strong interleaflet coupling and $16 - 48$ for unconnected bilayers[69]. The upper limit $b = 48$ is for bilayers where the two leaflets can freely slide past each other[68,70]. Experimental systems give $b$ in the range $20 - 30$.[68] At least two explicit solvent simulations gave $b = 48$.[70,71] Our estimates for well-equilibrated SiMPLISTIC bilayers gave $b = 31.7$ at $kT = 1.20\varepsilon_0$ and $b = 45.5$ at $kT = \varepsilon_0$. These values indicate that there is no significant interleaflet coupling for SiMPLISTIC bilayers. This is expected because we eliminated a key source for such coupling[72], viz. lipid interdigitation between the opposing leaflets, by design (Sec. II.A.4 and IV.A.3).

### B. Hydrophobic Mismatch

Hydrophobic mismatch is a key player in the physics of multicomponent lipid membranes if the constituent lipids differ in length due to a difference in either chain length or chain saturation[73–75]. Such difference in length between neighboring lipids in a leaflet exposes some of the hydrophobic parts of the longer lipid(s) to the surrounding water, which is unfavorable. To avoid this 'hydrophobic mismatch', lipids prefer to be surrounded by lipids with similar length, which may lead to a demixing of dissimilar lipids in the multicomponent system[76]. In other words, hydrophobic mismatch generates a line tension that, if large enough, may cause phase separation through domain formation[75,77,78].

SiMPLISTIC, as defined so far (Sec. II.C and III.C), can only model systems where all lipids have the same length ($\sigma_e$ in Sec. II.B.1). In order to model multicomponent membranes with hydrophobic mismatch, it therefore needs to handle interactions between lipids with different lengths. We extend SiMPLISTIC in the following by defining just such an interaction and test our model with simulations.

Let us take two types of SiMPLISTIC lipids (A and B) that differ only in their spheroidal lengths, $\sigma_e^A$ and $\sigma_e^B$. The contact distance between two such BP spheroids of unequal lengths had been worked out by Cleaver et al.[79] Accordingly, the contact distance $\sigma(\hat{\mathbf{r}}_{ij}, \hat{\mathbf{u}}_i, \hat{\mathbf{u}}_j)$ between a lipid $i$ of type A and another lipid $j$ of type B is given as

$$\sigma = \sigma_0 \left[ 1 - \left\{ \frac{\chi_A \left( \hat{\mathbf{r}}_{ij} \cdot \hat{\mathbf{u}}_i \right)^2 + \chi_B \left( \hat{\mathbf{r}}_{ji} \cdot \hat{\mathbf{u}}_j \right)^2 + 2 \chi_A \chi_B \left( \hat{\mathbf{r}}_{ij} \cdot \hat{\mathbf{u}}_i \right) \left( \hat{\mathbf{r}}_{ji} \cdot \hat{\mathbf{u}}_j \right) \left( \hat{\mathbf{u}}_i \cdot \hat{\mathbf{u}}_j \right)}{1 - \chi_A \chi_B \left( \hat{\mathbf{u}}_i \cdot \hat{\mathbf{u}}_j \right)^2} \right\} \right]^{-1/2}$$

$$(14)$$

where $\chi_A$ and $\chi_B$ denote the anisotropy parameters of $i$ and $j$ respectively, as defined in Sec. II.B.1. Note that for $i$ and $j$ lipids having identical lengths, $\chi_A = \chi_B = \chi$, which reduces $\sigma$ in Eq. (14) to $\sigma_{BP}$ in Eq. (5). With the contact distance ($\sigma$) obtained as above, we can now choose Eq. (6) as the steric part of the potential between $i$ and $j$, with $R$ now defined as $R = (r - \sigma + \sigma_0)/\sigma_0$.

Hydrophobic mismatch driven demixing, interpreted as a tendency of lipids of similar length to cluster together within a leaflet, can be modeled through the mechanism of like lipids interacting more strongly than unlike ones. This is most easily achieved by simply reducing the strength of the A-B interaction compared to that of the A-A and B-B interactions. Care must however be taken to do this for lipids within the same leaflet only. Otherwise, the weaker attraction between unlike lipids from opposing leaflets would lead to a direct interleaflet coupling manifested as domain registration between the leaflets. This is unphysical because hydrophobic mismatch, in reality, destabilizes registration rather than favoring it. For further illustration of this last point the reader is referred to Ref. [77] and the Fig. 2 and its corresponding discussion in Ref. [80].

In view of the above, let us define the energy scale (viz. $\varepsilon_0$ as defined in Sec. II.B) for the different pair interactions as

$$\varepsilon_0^{AA} = \varepsilon_0^{BB} = \varepsilon_0$$
$$\varepsilon_0^{AB} = \left[ 1 - p(\sigma_e^A, \sigma_e^A) \right] \varepsilon_0$$
where
$$p(\sigma_e^A, \sigma_e^B) > 0, \text{ for } \sigma_e^A \neq \sigma_e^B \text{ within same leaflet}$$
$$= 0, \text{ otherwise}$$

$$(15)$$

We, therefore, need to define $p(\sigma_e^A, \sigma_e^B)$ now. To this end, note that $p$ can be interpreted as a dimensionless penalty factor which determines the loss of binding energy due to lipid length mismatch. This loss of binding energy is what gives rise to the line tension ($\lambda$) between two neighboring domains of unlike lipids within a leaflet, and is directly proportional to it[75]. The line tension between domains of A and B, therefore, is directly proportional to the penalty factor, i.e. $\lambda \propto p$. Kuzmin et al.[81], in Eq. 17 of their paper, have derived a quadratic dependence of the line tension on $(\delta/h_0)$, where $\delta$ is the amount of hydrophobic mismatch and $h_0$ is the average thickness of the monolayer inside the two domains. This dependence is exact if the two domains have the same spontaneous curvature, which is a condition that applies well for flat membranes composed of lamellar lipids. In our case, $\delta \approx |\sigma_e^A - \sigma_e^B|$ and $h_0 \approx (\sigma_e^A + \sigma_e^B)/2$. Because $p$ scales with $\lambda$, and $\lambda$ scales with $(\delta/h_0)^2$ as discussed above, we may define $p$ as

$$p(\sigma_e^A, \sigma_e^B) = \nu \left( \sigma_e^A - \sigma_e^B \right)^2 \bigg/ \left( \frac{\sigma_e^A + \sigma_e^B}{2} \right)^2,$$

$$(16)$$

where $\nu$ is a positive, floating-point parameter that can be used to tune the line tension.

The interaction potential for a pair of A and B lipids is therefore given by Eq. (8) with the replacements $\sigma_{BP} \to \sigma$ and $\varepsilon_0 \to \varepsilon_0^{AB}$, where $\sigma$ and $\varepsilon_0^{AB}$ are given by Eq.s (14) and (15) respectively. Note that, with each lipid residing in either of the two leaflets of a flat membrane, the directed axes (viz. $\hat{\mathbf{u}}$) of a pair of lipids form an acute angle if the lipids are in the same leaflet and obtuse angle otherwise. Therefore, the condition for a pair of A and B lipids being in the same leaflet, and hence interacting with reduced strength, is simply $\hat{\mathbf{u}}_A \cdot \hat{\mathbf{u}}_B > 0$.

To test this model out, we simulated two-component membranes consisting of A and B lipids with $\sigma_e^A = 3\sigma_0$ and $\sigma_e^B = 4\sigma_0$, at $kT = \varepsilon_0$ and vanishing surface tension, for different values of $\nu$. Starting from a randomly mixed state, no demixing was apparent for $\nu = 1, 2$ even after long runs $\sim 10^5$ MD steps. This suggests that the corresponding line tensions for these smaller values of $\nu$ are not large enough to outweigh the entropic preference towards well-mixed states at $kT = \varepsilon_0$. With the increased line tension for $\nu \geq 3$, however, clusters of like lipids appeared with excellent large scale domain formation at $\nu = 5$ (Figure 11). As desired, no domain registration was observed between the opposing leaflets of the bilayer (Figure 11). SiMPLISTIC therefore, with the inclusion of interaction between lipids of different lengths as described above, succeeds in modeling hydrophobic mismatch.

## V. CONCLUSION

### A. Highlights

This paper described a novel, coarse-grained model called 'SiMPLISTIC' for large scale, particle based simulations of lamellar and non-lamellar lipid systems in implicit water. In absence of explicit water, the hydrophobic effect and the hydration force are mimicked with lipid-lipid attraction and repulsion respectively. 'SiMPLISTIC' is acronym for 'Single-site Model with Pairwise interaction for Lipids in Implicit Solvent with Tuneable Intrinsic Curvature'. The most important features of SiMPLISTIC are listed below.

a) SiMPLISTIC lipids are single-site (Sec. II.A.2), and follow rigid body dynamics as suggested by previous empirical studies (Sec. II.A.1). The soft-core, anisotropic, pair potential (Sec. II.C) contains only three parameters that relate to the following— the lipid length ($\sigma_e$), the interaction range (*range*) and the packing parameter ($m$).

b) SiMPLISTIC is a phenomenological model (Sec. II.A.4) designed for speed (Sec. V.B) and simplicity (Sec. II.A.3). The anisotropic nature of the SiMPLISTIC potential gives it a unique ability to model the hydration force between lipid aggregates as soft repulsions for certain relative orientations.

c) The model lipids show rapid, unassisted self-assembly (Sec. III.A.2). The free parameter $m$ determines the curvature of the self-assembled phase (Sec. III.B). Smaller values of $m$ generate inverted micelles and larger shapes, bilayers (Figure 6).

d) SiMPLISTIC includes mixing rules (Sec. III.C) for modeling mixtures of lipids with different packing parameters.

e) Model membrane simulations with bilayer forming SiMPLISTIC lipids show fluid-gel transition (Sec. IV.A.1) and yield experimentally relevant values for the bending stiffness in the fluid phase (Sec. IV.A.4). The bilayers are tilt-free (Sec. IV.A.2) and do not show interdigitation (IV.A.3). SiMPLISTIC membranes also give satisfactory elastic ratios (derived from the area compressibility modulus) that suggest the two leaflets of a SiMPLISTIC bilayer are not coupled strongly (Sec. IV.A.5).

f) A simple extension of SiMPLISTIC to model hydrophobic

mismatch (Sec. IV.B) successfully generates demixing through phase separation in multicomponent model membranes (Figure 11).

### B. Benchmarks

SiMPLISTIC is designed for speed. Here we report some crude benchmarks for SiMPLISTIC and compare it with some other relevant models.

The execution time of an optimized code is not a very meaningful metric for the underlying model's computational efficiency. This is because, apart from the platform, CPU time depends on the optimizations performed within the source code as well as during compilation, both of which are subject to variation from one implementation to another. Benchmarks under non-optimal conditions, on the other hand, are more useful because they represent the least performance expected from a model running on any given processor. With this in mind, we benchmarked SiMPLISTIC as follows.

The test code, implemented in Fortran 90 with double precision for the floating-points, consisted of just the SiMPLISTIC force routines and the Nosé-Hoover NVT integrator. To eliminate I/O calls during run, no configuration dumping was done. No parallelization or SIMD vectorization and no domain decomposition or neighbor list were applied – hence, the force routines enumerated every possible pair of lipids serially in the least efficient way. The compilation was done using gfortran with no optimization. The simulated system was the same as reported in Sec. III.A.1, with $m = 1$ and starting from a self-assembled phase of closely packed inverted micelles at $kT = \varepsilon_0$.

Although the code ran serially on a single core, no core affinity or task priority was set. Performance was measured using the simulation tracking tool "progrep".[82]

Running this test code for several times on a laptop with an Intel i3 processor (clock speed 2.13 GHz) and 32-bit GNU/Linux, we consistently obtained CPU times of 2.05 minutes per 1000 MD steps. We also timed the spontaneous self-assembly of the same system (i.e. 1600 lipids) from a disordered isotropic phase at 12.10 minutes (for 6000 MD steps). Needless to say, standard implementations employing neighbor lists, compiler optimizations, multicore parallelization, SIMD vectorization and GPU accelerations, will improve the execution time manifold. It may also be pointed out that the computation of $\left| \hat{\mathbf{u}}_i \cdot \hat{\mathbf{u}}_j \right|^m$

[Sec. II.B.3 and Eq. (8)] for floating point $m$ has a significant overhead in SiMPLISTIC. This is because such floating-point powers are computed using logarithms followed by exponentials – both of which operations are quite expensive. For integer $m$, however, $\left| \hat{\mathbf{u}}_i \cdot \hat{\mathbf{u}}_j \right|^m$ may be computed much faster because odd and even powers are best computed using a few (recursive) multiplications only.

To compare the computational performance of SiMPLISTIC to that of a standard single-site, rigid model with a similar anisotropic pair potential (viz. Gay-Berne[42]), and also a multi-site, bead-spring model for lipids in implicit solvent (viz. the Cooke-Kremer-Deserno model[30]), consider the following. SiMPLISTIC contains less floating point operations (flops) than a switched Gay-Berne (GB) potential and therefore, should be faster than the GB benchmarks on any given platform. SiMPLISTIC is also faster than a single-site amphiphile model previously published by us[35], again by virtue of requiring fewer flops. Being single-site, this amphiphile model was found to be more than 1.6 times faster than a rigid version of the multi-site Cooke-Kremer-Deserno model, which, in turn, is naturally faster than the original flexible (bead-spring) version[30]. SiMPLISTIC, therefore, should be more efficient than the Cooke-Kremer-Deserno model for large scale phenomena that are not significantly sensitive to lipid chain flexibility.

### C. Scope

Being molecular-level and rigid, SiMPLISTIC is targeted specifically towards supramolecular, lipid-based systems involving macro- and mesoscale phenomena not sensitive to lipid chain flexibility. In view of its computational economy and versatility (Sec.

V.A), SiMPLISTIC should be able to serve as an excellent model for large-scale lipid simulations in academic and industrial research. Because of its conceptual simplicity and effectiveness, SiMPLISTIC may also be used for educational purposes— e.g. for hands-on learning about concepts such as self-assembly, lipid polymorphism, biomembrane organization, domain formation etc., through lightweight, interactive simulation software.

To inspire future applications, some specific use cases for SiMPLISTIC are suggested below.

- Studying the discontinuous cubic phase (Fd3m) formed by inverted micelles.

- Studying elasticity, lipid diffusion and phase-separation behavior in multicomponent membranes as functions of the membrane constitution.

- Studying vesicles consisting of lamellar lipids on the outer surface and non-lamellar lipids on the inner. Hydrophobic mismatch driven demixing may induce budding in multicomponent vesicles.

- Modeling cholesterol as a non-lamellar lipid present in membranes composed of lamellar lipids. This, linked with SiMPLISTIC's ability to model hydrophobic mismatch and form lipid domains, might make studies of lipid rafts (or $L_d$-$L_o$ phase separation) relevant.

- Studying tension induced membrane rupture or pore formation.

- Modeling symmetric (asymmetric) bolalipids as dimers composed of two SiMPLISTIC lipids with identical (different) $m$, in the spirit of Ref. 36.

- Modeling lipid nanoparticles, liposomes and cubosomes for therapeutic applications such as drug and gene delivery.

- Studying membrane permeability by probing the permeation of simple, sterically interacting, Brownian particles through SiMPLISTIC membranes.

- Interactive molecular dynamics.

### D. Future Directions

#### 1. Two more phases with negative curvature

Although the free parameter $m$ tunes the spontaneous curvature continuously from inverted micelles to planar bilayers, it is curious that no inverted hexagonal ($H_{II}$) or bicontinuous/bilayer cubic ($Q_{II}^B$) phase was obtained through self-assembly in our simulations. We suspect that a cubic simulation box with periodic boundary conditions, as employed by us, cannot form the $H_{II}$ phase due to the lack of hexagonal symmetry. On the other hand, because it involves curved bilayers, the $Q_{II}^B$ phase would probably require a much larger system size. We hope to address these issues in future work(s).

#### 2. Possible extensions of SiMPLISTIC

SiMPLISTIC, as defined in this paper, is useful for simulating systems consisting of uncharged lipids only. Extending SiMPLISTIC to include interactions with purely hydrophilic and hydrophobic particles, which may be in the form of BP spheroids or spheres, will significantly increase its scope. Such an extension, for example, might be useful in modeling lipid-protein interactions and studying membrane permeability or leakage of cargo from cubosomes and liposomes.

In addition to this, defining electrostatic charges on SiMPLISTIC lipids and other particles will make the model even more effective. Such an extension might enable probing important large-scale phenomena such as lipid-DNA complex formation in the presence of counter-ions, fusion of oppositely charged vesicles etc.

### 3. An open question

SiMPLISTIC may be seen as a proof of concept, which demonstrates that a single-site, rigid model designed using some simple guidelines (Sec. II.A) can spontaneously self-assemble into fluid phases and generate elastic properties consistent with experiments and other simulations. Moreover, it shows how the intrinsic curvature of the self-assembled phase (or equivalently, the molecular packing parameter[58]) may be tuned with only a single parameter. Our previous single-site and rigid amphiphile model[35] was also capable of tuning the curvature with a single parameter. However, that model gave phases with positive curvatures, whereas SiMPLISTIC gives phases with negative curvatures. To resolve this dichotomy, it is natural to imagine a single-site, rigid model with simple, anisotropic, pairwise interaction that will self-assemble into phases with any possible curvature – positive or negative – depending on the value of a single tuneable parameter. Apart from its sheer elegance, such a minimalistic, molecular level model will also be extremely useful, as amphiphiles with any packing parameter may be modeled with it, including mixtures of molecules with different effective shapes (packing parameters). Whether such an all-encompassing yet computationally economic model can be designed is, however, an open question.

### ACKNOWLEDGMENTS

Somajit Dey is funded by the Council of Scientific & Industrial Research (CSIR), India, through a Senior Research Fellowship [Grant No. 09/028(0960)/2015-EMR-I]. Dey thanks Dr. Sayantani Das for improving some of the illustrations.

### AUTHOR INFORMATION

*Somajit Dey* — Email: sdphys_rs@caluniv.ac.in; Current Address: Department of Physics, University of Calcutta, 92, A.P.C. Road, Kolkata-700009, India;  https://orcid.org/0000-0002-6102-9777

*Jayashree Saha* — Email: jsphy@caluniv.ac.in; Current Address: Department of Physics, University of Calcutta, 92, A.P.C. Road, Kolkata-700009, India;

### REFERENCES

(1)  Mouritsen, O. G.; Bagatolli, L. A. *LIFE- AS A MATTER OF FAT Lipids in a Membrane Biophysics Perspective*; The Frontiers Collection; Springer International Publishing: Cham, 2016. https://doi.org/10.1007/978-3-319-22614-9.

(2)  Israelachvili, J. N.; Mitchell, D. J.; Ninham, B. W. Theory of Self-Assembly of Hydrocarbon Amphiphiles into Micelles and Bilayers. *J. Chem. Soc. Faraday Trans. 2 Mol. Chem. Phys.* **1976**, *72*, 1525–1568. https://doi.org/10.1039/F29767201525.

(3)  Lombardo, D.; Kiselev, M. A.; Magazù, S.; Calandra, P. Amphiphiles Self-Assembly: Basic Concepts and Future Perspectives of Supramolecular Approaches. *Adv. Condens. Matter Phys.* **2015**, *2015*. https://doi.org/10.1155/2015/151683.

(4)  Luzzati, V.; Delacroix, H.; Gulik, A.; Gulik-Krzywicki, T.; Mariani, P.; Vargas, R. The Cubic Phases of Lipids. *Stud. Surf. Sci. Catal.* **2004**, *148* (C), 17–40. https://doi.org/10.1016/S0167-2991(04)80191-5.

(5)  Henriksen, J. R.; Andresen, T. L.; Feldborg, L. N.; Duelund, L.; Ipsen, J. H. Understanding Detergent Effects on Lipid Membranes: A Model Study of Lysolipids. *Biophys. J.* **2010**, *98* (9), 2199–2205. https://doi.org/10.1016/j.bpj.2010.01.037.

(6)  Helfrich, W. Lyotropic Lamellar Phases. *J. Phys. Condens. Matter* **1994**, *6* (23A). https://doi.org/10.1088/0953-8984/6/23A/009.

(7)  Damodaran, K. V.; Merz, K. M. Computer Simulation of Lipid Systems. In *Reviews in Computational Chemistry V*; Lipkowitz, K. B., Boyd, D. B., Eds.; VCH Publishers, Inc, 2007;  pp  269–298. https://doi.org/10.1002/9780470125823.ch5.

(8)  Tenchov, B.; Koynova, R. Cubic Phases in Membrane Lipids. *Eur. Biophys. J.* **2012**, *41* (10), 841–850. https://doi.org/10.1007/s00249-012-0819-3.

(9)  Van Meer, G.; De Kroon, A. I. P. M. Lipid Map of the Mammalian Cell. *J. Cell Sci.* **2011**, *124* (1), 5–8. https://doi.org/10.1242/jcs.071233.

(10)  van Meer, G.; Voelker, D. R.; Feigenson, G. W. Membrane Lipids: Where They Are and How They Behave. *Nat. Rev. Mol. Cell Biol.* **2008**, *9* (2), 112–124. https://doi.org/10.1038/nrm2330.

(11)  Basu Ball, W.; Neff, J. K.; Gohil, V. M. The Role of Nonbilayer Phospholipids in Mitochondrial Structure and Function. *FEBS Lett.* **2018**, *592* (8), 1273–1290. https://doi.org/10.1002/1873-3468.12887.

(12)  Azmi, I. D. M.; Moghimi, S. M.; Yaghmur, A. Cubosomes and Hexosomes as Versatile Platforms for Drug Delivery. *Ther. Deliv.* **2015**, *6* (12), 1347–1364. https://doi.org/10.4155/tde.15.81.

(13)  Koynova, R.; Tenchov, B.; Macdonald, R. C. Nonlamellar Phases in Cationic Phospholipids, Relevance to Drug and Gene Delivery. *ACS Biomater. Sci. Eng.* **2015**, *1* (3), 130–138. https://doi.org/10.1021/ab500142w.

(14)  Kulkarni, C. V. Lipid Self-Assemblies and Nanostructured Emulsions for Cosmetic Formulations. *Cosmetics* **2016**, *3* (4). https://doi.org/10.3390/cosmetics3040037.

(15)  Akbarzadeh, A.; Rezaei-Sadabady, R.; Davaran, S.; Joo, S. W.; Zarghami, N.; Hanifehpour, Y.; Samiei, M.; Kouhi, M.; Nejati-Koshki, K. Liposome: Classification, Preparation, and Applications. *Nanoscale Res. Lett.* **2013**, *8* (1), 1. https://doi.org/10.1186/1556-276X-8-102.

(16)  Eeman, M.; Deleu, M. From Biological Membranes to Biomimetic Model Membranes. *Biotechnol. Agron. Soc. Environ.* **2010**, *14* (4), 719–736.

(17)  Daraee, H.; Etemadi, A.; Kouhi, M.; Alimirzalu, S.; Akbarzadeh, A. Application of Liposomes in Medicine and Drug Delivery. *Artif. Cells, Nanomedicine Biotechnol.* **2016**, *44* (1), 381–391. https://doi.org/10.3109/21691401.2014.953633.

(18)  Balazs, D. A.; Godbey, W. Liposomes for Use in Gene Delivery. *J. Drug Deliv.* **2011**, *2011*, 1–12. https://doi.org/10.1155/2011/326497.

(19)  Handjani-vila, R. M.; Ribier, A.; Rondot, B.; Vanlerberghie, G. Dispersions of Lamellar Phases of Non-Ionic Lipids in Cosmetic Products. *Int. J. Cosmet. Sci.* **1979**, *1* (5), 303–314. https://doi.org/10.1111/j.1467-2494.1979.tb00224.x.

(20)  Vattulainen, I.; Rog, T. Lipid Simulations: A Perspective on Lipids in Action. *Cold Spring Harb. Perspect. Biol.* **2011**, *3* (4), 1–13. https://doi.org/10.1101/cshperspect.a004655.

(21)  Klein, M. L.; Shinoda, W. Large-Scale Molecular Dynamics Simulations of Self-Assembling Systems. *Science (80-. ).* **2008**, *321* (5890), 798–800. https://doi.org/10.1126/science.1157834.

(22)  Allen, D. T.; Lorenz, C. D. Molecular Scale Simulations of the Self-Assembly of Amphiphilic Molecules: Current State-of-the-Art and Future Directions. *J. Self-Assembly Mol. Electron.* **2015**, *2015* (1), 01–38. https://doi.org/10.13052/jsame2245-4551.2015006.

(23)  *Coarse-Graining of Condensed Phase and Biomolecular Systems*; Voth, G. A., Ed.; CRC Press: Boca Raton, 2008. https://doi.org/10.1201/9781420059564.

(24)  Brannigan, G.; Lin, L. C. L.; Brown, F. L. H. Implicit Solvent Simulation Models for Biomembranes. *Eur. Biophys. J.* **2006**, *35* (2), 104–124. https://doi.org/10.1007/s00249-005-0013-y.

(25)  Drouffe, J. M.; Maggs, A. C.; Leibler, S. Computer Simulations of Self-Assembled Membranes. *Science (80-. ).* **1991**, *254* (5036), 1353–1356. https://doi.org/10.1126/science.1962193.

(26)  Noguchi, H.; Takasu, M. Self-Assembly of Amphiphiles into Vesicles: A Brownian Dynamics Simulation. *Phys. Rev. E - Stat. Physics, Plasmas, Fluids, Relat. Interdiscip. Top.* **2001**, *64* (4), 7. https://doi.org/10.1103/PhysRevE.64.041913.

(27)  Farago, O. "Water-Free" Computer Model for Fluid Bilayer

Membranes. *J. Chem. Phys.* **2003**, *119* (1), 596–605. https://doi.org/10.1063/1.1578612.

(28) Brannigan, G.; Brown, F. L. H. Solvent-Free Simulations of Fluid Membrane Bilayers. *J. Chem. Phys.* **2004**, *120* (2), 1059–1071. https://doi.org/10.1063/1.1625913.

(29) Wang, Z. J.; Frenkel, D. Modeling Flexible Amphiphilic Bilayers: A Solvent-Free off-Lattice Monte Carlo Study. *J. Chem. Phys.* **2005**, *122* (23). https://doi.org/10.1063/1.1927509.

(30) Cooke, I. R.; Kremer, K.; Deserno, M. Tunable Generic Model for Fluid Bilayer Membranes. *Phys. Rev. E - Stat. Nonlinear, Soft Matter Phys.* **2005**, *72* (1), 2–5. https://doi.org/10.1103/PhysRevE.72.011506.

(31) Cooke, I. R.; Deserno, M. Solvent-Free Model for Self-Assembling Fluid Bilayer Membranes: Stabilization of the Fluid Phase Based on Broad Attractive Tail Potentials. *J. Chem. Phys.* **2005**, *123* (22). https://doi.org/10.1063/1.2135785.

(32) Noguchi, H. Solvent-Free Coarse-Grained Lipid Model for Large-Scale Simulations. *J. Chem. Phys.* **2011**, *134* (5), 0–12. https://doi.org/10.1063/1.3541246.

(33) Rand, R. P.; Parsegian, V. A. Hydration Forces between Phospholipid Bilayers. *Biochim. Biophys. Acta - Rev. Biomembr.* **1989**, *988* (3), 351–376. https://doi.org/10.1016/0304-4157(89)90010-5.

(34) Leckband, D.; Israelachvili, J. *Intermolecular Forces in Biology*; 2001; Vol. 34. https://doi.org/10.1017/s0033583501003687.

(35) Dey, S.; Saha, J. Solvent-Free, Molecular-Level Modeling of Self-Assembling Amphiphiles in Water. *Phys. Rev. E* **2017**, *95* (2), 023315. https://doi.org/10.1103/PhysRevE.95.023315.

(36) Dey, S.; Saha, J. Minimal Coarse-Grained Modeling toward Implicit Solvent Simulation of Generic Bolaamphiphiles. *J. Phys. Chem. B* **2020**, *124* (14), 2938–2949. https://doi.org/10.1021/acs.jpcb.0c00734.

(37) Klauda, J. B.; Eldho, N. V.; Gawrisch, K.; Brooks, B. R.; Pastor, R. W. Collective and Noncollective Models of NMR Relaxation in Lipid Vesicles and Multilayers †. *J. Phys. Chem. B* **2008**, *112* (19), 5924–5929. https://doi.org/10.1021/jp075641w.

(38) Pastor, R. W.; Venable, R. M.; Karplus, M.; Szabo, A. A Simulation Based Model of NMR T 1 Relaxation in Lipid Bilayer Vesicles. *J. Chem. Phys.* **1988**, *89* (2), 1128–1140. https://doi.org/10.1063/1.455219.

(39) Pastor, R. W.; Venable, R. M.; Feller, S. E. Lipid Bilayers, NMR Relaxation, and Computer Simulations. *Acc. Chem. Res.* **2002**, *35* (6), 438–446. https://doi.org/10.1021/ar0100529.

(40) Sun, X.; Gezelter, J. D. Dipolar Ordering in the Ripple Phases of Molecular-Scale Models of Lipid Membranes. *J. Phys. Chem. B* **2008**, *112* (7), 1968–1975. https://doi.org/10.1021/jp0762020.

(41) Marsh, D. Molecular Motion in Phospholipid Bilayers in the Gel Phase: Long Axis Rotation. *Biochemistry* **1980**, *19* (8), 1632–1637. https://doi.org/10.1021/bi00549a017.

(42) Gay, J. G.; Berne, B. J. Modification of the Overlap Potential to Mimic a Linear Site-Site Potential. *J. Chem. Phys.* **1981**, *74* (6), 3316–3319. https://doi.org/10.1063/1.441483.

(43) Bates, M. A.; Luckhurst, G. R. Computer Simulation of Liquid Crystal Phases Formed by Gay-Berne Mesogens. In *Liquid Crystals I*; Springer Berlin Heidelberg: Heidelberg, 1999; Vol. 94, pp 65–137. https://doi.org/10.1007/3-540-68305-4_3.

(44) Wilson, M. R. Molecular Simulation of Liquid Crystals: Progress towards a Better Understanding of Bulk Structure and the Prediction of Material Properties. *Chem. Soc. Rev.* **2007**, *36* (12), 1881–1888. https://doi.org/10.1039/b612799c.

(45) Allen, M. P. Molecular Simulation of Liquid Crystals. *Mol. Phys.* **2019**, *117* (18), 2391–2417. https://doi.org/10.1080/00268976.2019.1612957.

(46) Weeks, J. D.; Chandler, D.; Andersen, H. C. Role of Repulsive Forces in Determining the Equilibrium Structure of Simple Liquids. *J. Chem. Phys.* **1971**, *54* (12), 5237–5247. https://doi.org/10.1063/1.1674820.

(47) Heyes, D. M.; Okumura, H. Some Physical Properties of the Weeks–Chandler–Andersen Fluid. *Mol. Simul.* **2006**, *32* (1), 45–50. https://doi.org/10.1080/08927020500529442.

(48) Berne, B. J.; Pechukas, P. Gaussian Model Potentials for Molecular Interactions. *J. Chem. Phys.* **1972**, *56* (8), 4213–4216. https://doi.org/10.1063/1.1677837.

(49) Vega, C.; Lago, S. A Fast Algorithm to Evaluate the Shortest Distance between Rods. *Comput. Chem.* **1994**, *18* (1), 55–59. https://doi.org/10.1016/0097-8485(94)80023-5.

(50) Allen, M. P.; Germano, G. Expressions for Forces and Torques in Molecular Simulations Using Rigid Bodies. *Mol. Phys.* **2006**, *104* (20–21), 3225–3235. https://doi.org/10.1080/00268970601075238.

(51) Arnarez, C.; Uusitalo, J. J.; Masman, M. F.; Ingólfsson, H. I.; De Jong, D. H.; Melo, M. N.; Periole, X.; De Vries, A. H.; Marrink, S. J. Dry Martini, a Coarse-Grained Force Field for Lipid Membrane Simulations with Implicit Solvent. *J. Chem. Theory Comput.* **2015**, *11* (1), 260–275. https://doi.org/10.1021/ct500477k.

(52) Brannigan, G.; Philips, P. F.; Brown, F. L. H. Flexible Lipid Bilayers in Implicit Solvent. *Phys. Rev. E - Stat. Nonlinear, Soft Matter Phys.* **2005**, *72* (1), 4–7. https://doi.org/10.1103/PhysRevE.72.011915.

(53) Hoover, W. G. Canonical Dynamics: Equilibrium Phase-Space Distributions. *Phys. Rev. A* **1985**, *31* (3), 1695–1697. https://doi.org/10.1103/PhysRevA.31.1695.

(54) Nosé, S. An Extension of the Canonical Ensemble Molecular Dynamics Method. *Mol. Phys.* **1986**, *57* (1), 187–191. https://doi.org/10.1080/00268978600100141.

(55) Holian, B. L.; Voter, A. F.; Ravelo, R. Thermostatted Molecular Dynamics: How to Avoid the Toda Demon Hidden in Nosé-Hoover Dynamics. *Phys. Rev. E* **1995**, *52* (3), 2338–2347. https://doi.org/10.1103/PhysRevE.52.2338.

(56) Dey, S. Time-Reversible, Symplectic, Angular Velocity Based Integrator for Rigid Linear Molecules. **2018**, arXiv:1811.06450.

(57) Dey, S. Minimal Modification to Nose-Hoover Barostat Enables Correct NPT Sampling. **2020**, arXiv:2007.01838.

(58) Kunz, W.; Testard, F.; Zemb, T. Correspondence between Curvature, Packing Parameter, and Hydrophilic-Lipophilic Deviation Scales around the Phase-Inversion Temperature. *Langmuir* **2009**, *25* (1), 112–115. https://doi.org/10.1021/la8028879.

(59) Feller, S. E.; Pastor, R. W. Constant Surface Tension Simulations of Lipid Bilayers: The Sensitivity of Surface Areas and Compressibilities. *J. Chem. Phys.* **1999**, *111* (3), 1281–1287. https://doi.org/10.1063/1.479313.

(60) Jähnig, F. What Is the Surface Tension of a Lipid Bilayer Membrane? *Biophys. J.* **1996**, *71* (3), 1348–1349. https://doi.org/10.1016/S0006-3495(96)79336-0.

(61) Allen, M. P.; Tildesley, D. J. *Computer Simulation of Liquids*; Clarendon Press: Oxford, 1987. https://doi.org/10.1093/oso/9780198803195.001.0001.

(62) Frenkel, D.; Smit, B. *Understanding Molecular Simulation*; Academic Press, 2002.

(63) Dimova, R. Recent Developments in the Field of Bending Rigidity Measurements on Membranes. *Adv. Colloid Interface Sci.* **2014**, *208*, 225–234. https://doi.org/10.1016/j.cis.2014.03.003.

(64) Johner, N.; Harries, D.; Khelashvili, G. Curvature and Lipid Packing Modulate the Elastic Properties of Lipid Assemblies: Comparing HII and Lamellar Phases. *J. Phys. Chem. Lett.* **2014**, *5* (23), 4201–4206. https://doi.org/10.1021/jz5022284.

(65) Doktorova, M.; Harries, D.; Khelashvili, G. Determination of Bending Rigidity and Tilt Modulus of Lipid Membranes from

Real-Space Fluctuation Analysis of Molecular Dynamics Simulations. *Phys. Chem. Chem. Phys.* **2017**, *19* (25), 16806–16818. https://doi.org/10.1039/C7CP01921A.

(66) Johner, N.; Harries, D.; Khelashvili, G. Implementation of a Methodology for Determining Elastic Properties of Lipid Assemblies from Molecular Dynamics Simulations. *BMC Bioinformatics* **2016**, *17* (1), 1–11. https://doi.org/10.1186/s12859-016-1003-z.

(67) Sackmann, E. Physical Basis of Self-Organization and Function of Membranes: Physics of Vesicles. In *Handbook of Biological Physics*; Lipowsky, R., Sackmann, E., Eds.; 1995; Vol. 1, pp 213–304. https://doi.org/10.1016/S1383-8121(06)80022-9.

(68) Rawicz, W.; Olbrich, K. C.; McIntosh, T.; Needham, D.; Evans, E. Effect of Chain Length and Unsaturation on Elasticity of Lipid Bilayers. *Biophys. J.* **2000**, *79* (1), 328–339. https://doi.org/10.1016/S0006-3495(00)76295-3.

(69) Bloom, M.; Evans, E.; Mouritsen, O. G. Physical Properties of the Fluid Lipid-Bilayer Component of Cell Membranes: A Perspective. *Q. Rev. Biophys.* **1991**, *24* (3), 293–397. https://doi.org/10.1017/S0033583500003735.

(70) Goetz, R.; Gompper, G.; Lipowsky, R. Mobility and Elasticity of Self-Assembled Membranes. **1999**, 221–224.

(71) Marrink, S. J.; Mark, A. E. Effect of Undulations on Surface Tension in Simulated Bilayers. *J. Phys. Chem. B* **2001**, *105* (26), 6122–6127. https://doi.org/10.1021/jp0103474.

(72) May, S. Trans-Monolayer Coupling of Fluid Domains in Lipid Bilayers. *Soft Matter* **2009**, *5* (15), 3148–3156. https://doi.org/10.1039/b901647c.

(73) Risbo, J.; Sperotto, M. M.; Mouritsen, O. G. Theory of Phase Equilibria and Critical Mixing Points in Binary Lipid Bilayers. *J. Chem. Phys.* **1995**, *103* (9), 3643–3656. https://doi.org/10.1063/1.470041.

(74) Jørgensen, K.; Sperotto, M. M.; Mouritsen, O. G.; Ipsen, J. H.; Zuckermann, M. J. Phase Equilibria and Local Structure in Binary Lipid Bilayers. *BBA - Biomembr.* **1993**, *1152* (1), 135–145. https://doi.org/10.1016/0005-2736(93)90240-Z.

(75) Wallace, E. J.; Hooper, N. M.; Olmsted, P. D. Effect of Hydrophobic Mismatch on Phase Behavior of Lipid Membranes. *Biophys. J.* **2006**, *90* (11), 4104–4118. https://doi.org/10.1529/biophysj.105.062778.

(76) Knoll, W.; Schmidt, G.; Sackmann, E.; Ibel, K. Critical Demixing in Fluid Bilayers of Phospholipid Mixtures. A Neutron Diffraction Study. *J. Chem. Phys.* **1983**, *79* (7), 3439–3442. https://doi.org/10.1063/1.446193.

(77) Fowler, P. W.; Williamson, J. J.; Sansom, M. S. P.; Olmsted, P. D. Roles of Interleaflet Coupling and Hydrophobic Mismatch in Lipid Membrane Phase-Separation Kinetics. *J. Am. Chem. Soc.* **2016**, *138* (36), 11633–11642. https://doi.org/10.1021/jacs.6b04880.

(78) García-Sáez, A. J.; Chiantia, S.; Schwille, P. Effect of Line Tension on the Lateral Organization of Lipid Membranes. *J. Biol. Chem.* **2007**, *282* (46), 33537–33544. https://doi.org/10.1074/jbc.M706162200.

(79) Cleaver, D. J.; Care, C. M.; Allen, M. P.; Neal, M. P. Extension and Generalization of the Gay-Berne Potential. *Phys. Rev. E - Stat. Physics, Plasmas, Fluids, Relat. Interdiscip. Top.* **1996**, *54* (1), 559–567. https://doi.org/10.1103/PhysRevE.54.559.

(80) Fujimoto, T.; Parmryd, I. Interleaflet Coupling, Pinning, and Leaflet Asymmetry-Major Players in Plasma Membrane Nanodomain Formation. *Front. Cell Dev. Biol.* **2017**, *4* (JAN), 1–12. https://doi.org/10.3389/fcell.2016.00155.

(81) Kuzmin, P. I.; Akimov, S. A.; Chizmadzhev, Y. A.; Zimmerberg, J.; Cohen, F. S. Line Tension and Interaction Energies of Membrane Rafts Calculated from Lipid Splay and Tilt. *Biophys. J.* **2005**, *88* (2), 1120–1133. https://doi.org/10.1529/biophysj.104.048223.

(82) Dey, S. Progrep. https://doi.org/10.5281/zenodo.4294762.

(83) Gabriel, A. T.; Meyer, T.; Germano, G. Molecular Graphics of Convex Body Fluids. *J. Chem. Theory Comput.* **2008**, *4* (3), 468–476. https://doi.org/10.1021/ct700192z.

(84) Sodt, A. J.; Head-Gordon, T. An Implicit Solvent Coarse-Grained Lipid Model with Correct Stress Profile. *J. Chem. Phys.* **2010**, *132* (20), 1–8. https://doi.org/10.1063/1.3408285.

(85) Curtis, E. M.; Hall, C. K. Molecular Dynamics Simulations of DPPC Bilayers Using "LIME", a New Coarse-Grained Model. *J. Phys. Chem. B* **2013**, *117* (17), 5019–5030. https://doi.org/10.1021/jp309712b.

(86) Grime, J. M. A.; Madsen, J. J. Efficient Simulation of Tunable Lipid Assemblies Across Scales and Resolutions. **2019**, arXiv:1910.05362.

(87) Ionov, R.; Angelova, A. Swelling of Bilayer Lipid Membranes. *Thin Solid Films* **1996**, *284–285*, 809–812. https://doi.org/10.1016/S0040-6090(95)08452-5.

(88) Cooke, I. R.; Deserno, M. Coupling between Lipid Shape and Membrane Curvature. *Biophys. J.* **2006**, *91* (2), 487–495. https://doi.org/10.1529/biophysj.105.078683.

**FIGURES**

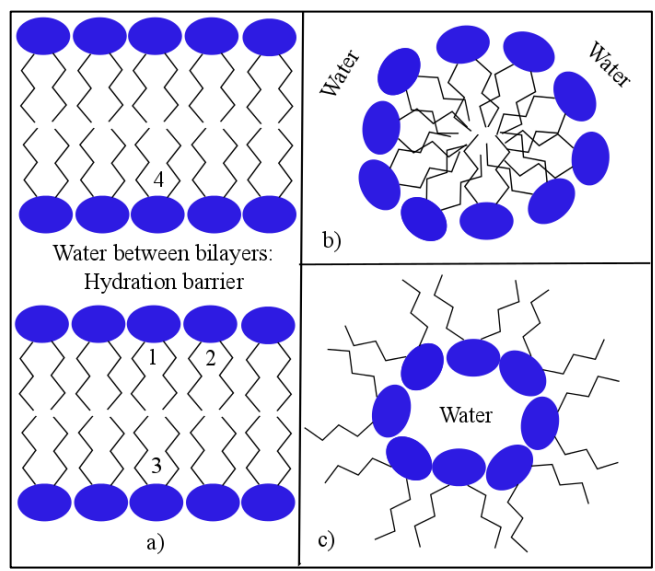

Figure 1. Schematic diagrams of lipid aggregates with different curvatures. Lipid cartoon: Blue – hydrophilic headgroup, Jagged lines – hydrophobic tails. a) Two hydrated bilayers separated by hydration force (No curvature); b) Cross-section of direct micelle (Positive curvature); c) Cross-section of inverted micelle (Negative curvature).

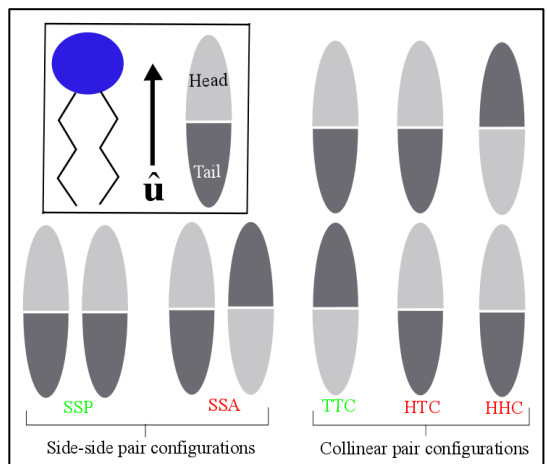

Figure 2. Inset: Lipid represented as a directed ellipsoid. Five canonical pair configurations— SSP: Side-Side Parallel, SSA: Side-Side Antiparallel, TTC: Tail-Tail Collinear, HTC: Head-Tail Collinear, HHC: Head-Head Collinear. Green ink denotes bound configuration (potential well-depth > 0); Red ink denotes unbound configuration (potential well-depth < 0).

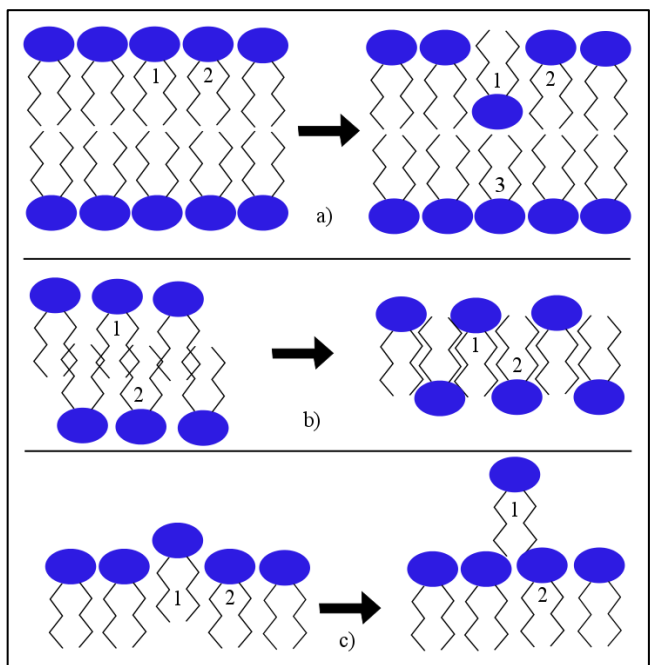

Figure 3. a) Lipid flip within a leaflet. The flipped lipid (1) forms SSA configuration with 2 and HTC with 3; b) Interdigitation between leaflets. Lipids 1 and 2 become SSA after interdigitation; c) Lipid protrusion from a leaflet. The protruded lipid (1) is almost HTC with 2.

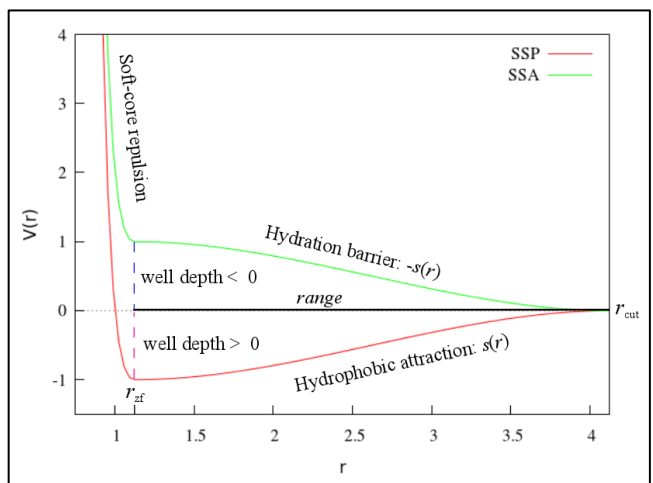

Figure 4. Potential energy between a pair of SiMPLISTIC lipids in SSP and SSA orientations as functions of their centre-centre distance, $r$. SSP must be stable and bound, hence the attractive tail of the potential. SSA must be unstable and unbound, hence the repulsive tail. Likewise, the TTC orientation has an attractive tail and each of the HTC and HHC orientations has a repulsive tail (not shown here).

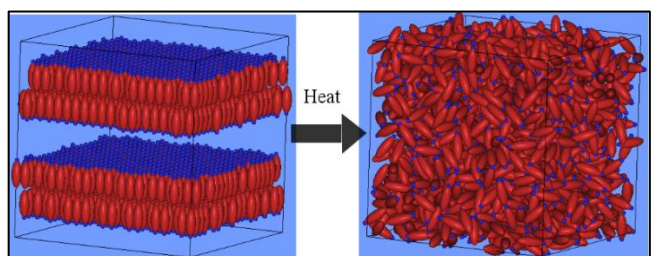

Figure 5. How the disordered, isotropic phase was prepared from preassembled lattice bilayers with sufficient interbilayer distance. SiMPLISTIC lipids are depicted as red spheroids with their head-ends marked with blue beads. The sky blue background implies the aqueous environment. This and other such renderings depicted in this work were performed using the molecular graphics software QMGA[83].

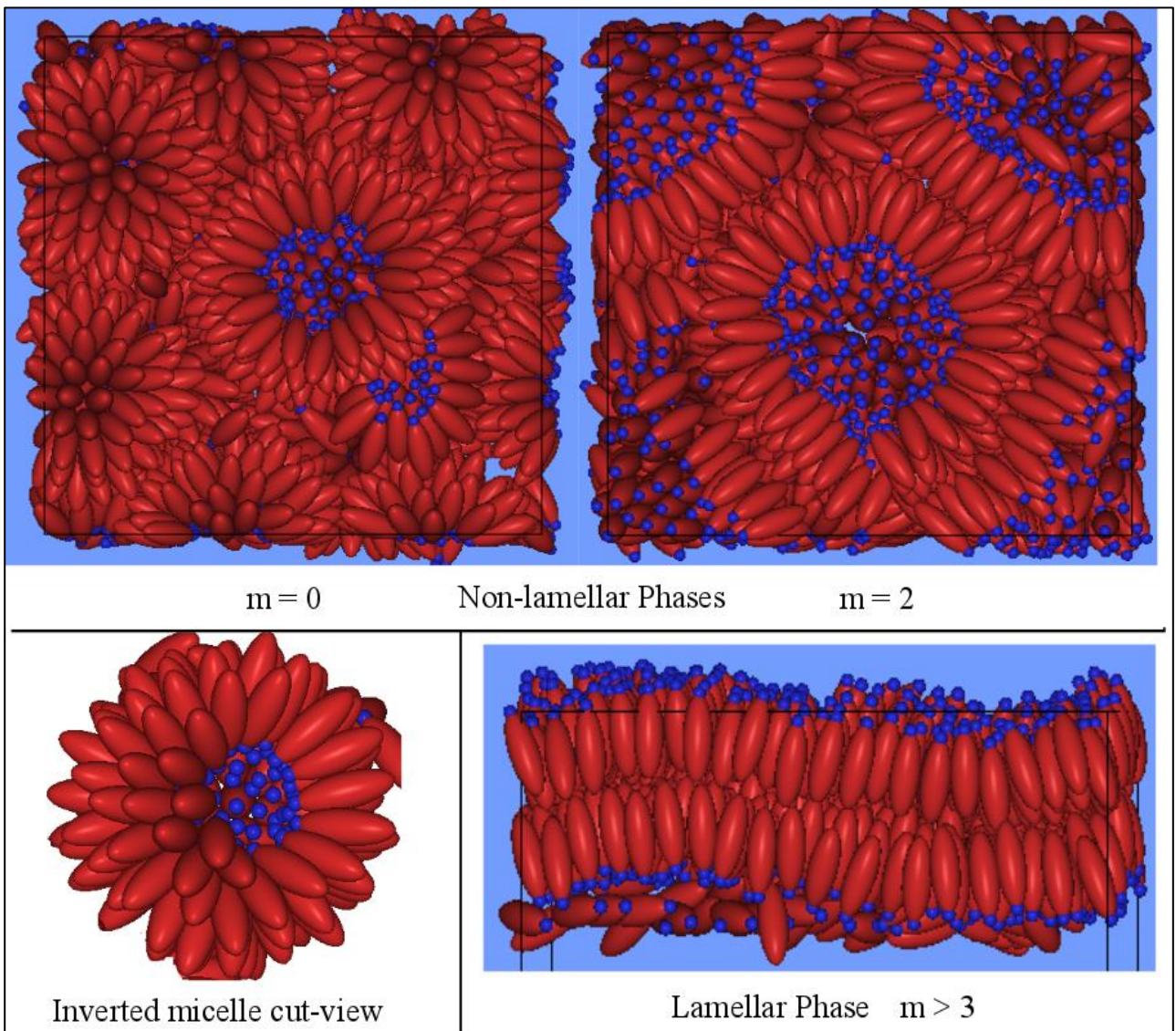

Figure 6. Self-assembled phases formed by SiMPLISTIC for different values of $m$. Lipids are depicted as red spheroids with their head-ends marked blue. The sky blue background implies the water environment. The non-lamellar phases are formed by closely packed inverted micelles (compare with the schematic cross-section in Figure 1). For enhanced clarity, a slice of an inverted micelle is also shown. The micellar radii are greater (and equivalently, the curvature is lesser) for $m = 2$ compared to $m = 0$. For $m > 3$ bilayers are obtained— a close-up of a box-spanning bilayer is shown [the stray lipids below the lower leaflet of the bilayer are actually part of a separate aggregate in the background (that we have sliced off for clarity) and do not belong to the bilayer in focus].

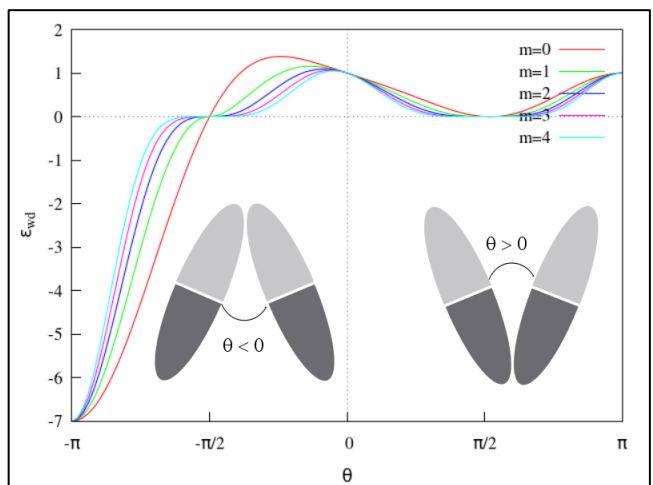

Figure 7. Well-depth $\varepsilon_{wd}$ of the SiMPLISTIC potential as a function of the lipid splay angle $\theta$ for different values of $m$. The peak shifts towards $\theta = 0$ for increasing $m$. Hence, increased $m$ leads to a preference for decreased splay. It is easy to see from the spheroid configurations that packing pairs of lipids with $\theta < 0$ would generate inverted micelles (Figure 1c). Hence, the effective molecular shape (packing parameter) of lipids preferring $\theta < 0$ is similar to an inverted (truncated) cone.

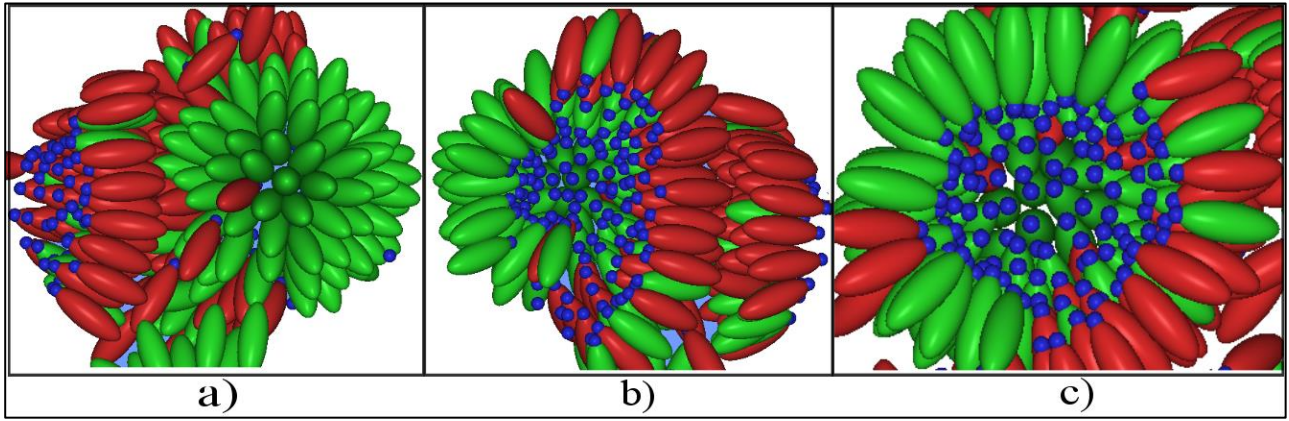

Figure 8. Close-ups of self-assembled phases generated by mixtures of two types of SiMPLISTIC lipids differing only in $m$. Red lipids have $m = 4$ and green ones have $m = 0$. Green lipids prefer inverted micelles (a), whereas red lipids prefer bilayer formation (b). Panel c shows the cross-section of an inverted micelle.

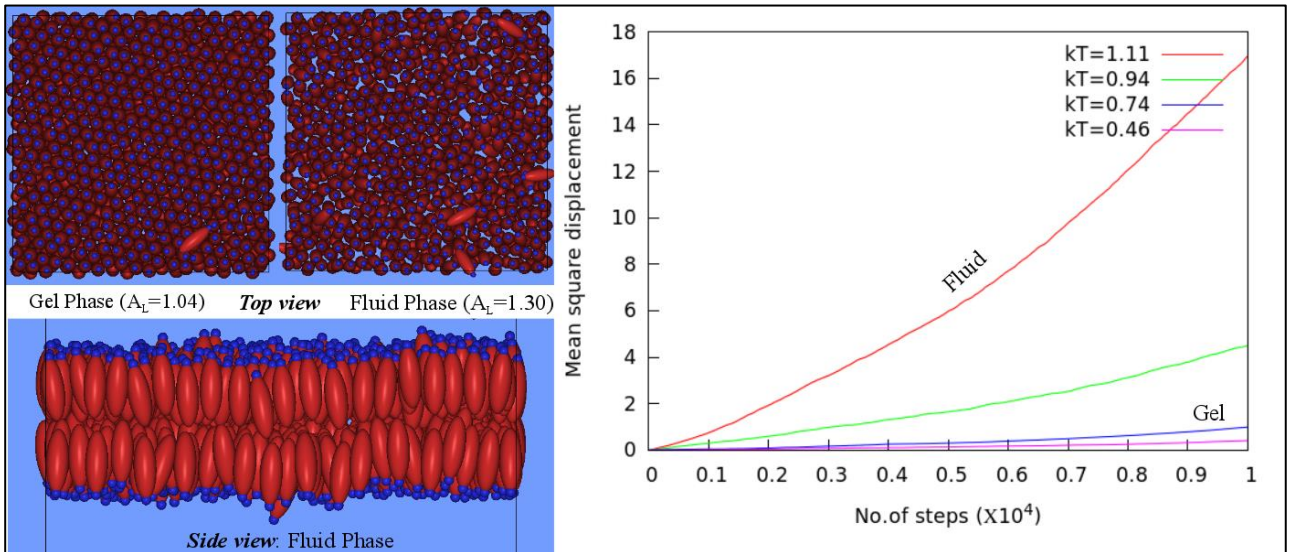

Figure 9. SiMPLISTIC shows fluid-gel transition in model membranes. The membrane phase can be deduced from the mean square displacements, and the order of lipid arrangement in the leaflets. Lipids in the fluid phase are much more diffusive but much less ordered and less densely packed (hence, the greater area per lipid, $A_L$). Note that no interdigitation or tilt is apparent in the side view. Units: $kT$ in $\varepsilon_0$ and $A_L$ in $\sigma_0^2$.

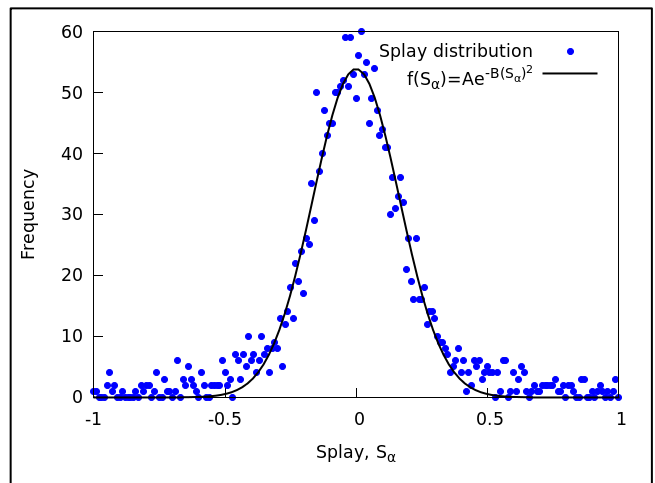

Figure 10. A Gaussian with zero mean fitted to the frequency distribution of splay obtained from a tensionless membrane at temperature, $kT = 1.11\varepsilon_0$. An estimate for the membrane bending modulus $\kappa_C$ can be obtained from the fit as $\kappa_C = 2BkT/A_L$, where $A_L$ is the area per lipid.

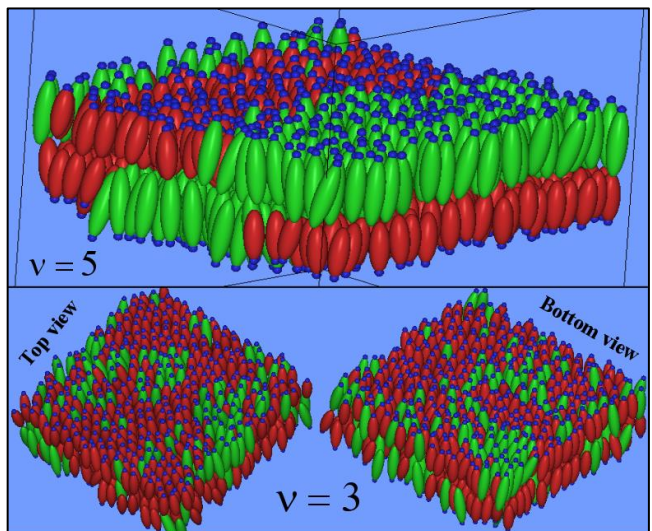

Figure 11. Hydrophobic mismatch driven demixing of SiMPLISTIC lipids at temperature, $kT = \varepsilon_0$. Red and green lipids differ only in their spheroidal lengths ($\sigma_e$). Red: $\sigma_e = 3\sigma_0$, Green: $\sigma_e = 4\sigma_0$. For $\nu = 3$, the number ratio of red to green is 3:1. For $\nu = 5$, the number ratio is 1:1. Although clustering of lipids of the same color in each leaflet is clearly visible for both values of $\nu$, a much cleaner phase separation occurs at $\nu = 5$ due to an increased line tension. Note that there is no domain registration between the leaflets.